\documentclass[twocolumn,aps,prc,superscriptaddress,longbibliography,showpacs,nofootinbib,floatfix]{revtex4-1}

\newcommand{\pt}{\mbox{$p_T$}\xspace}

\newcommand{\Np}{\mbox{$N_{\rm part}$}\xspace}

\newcommand{\sqsn}{\mbox{$\sqrt{s_{_{NN}}}$}\xspace}
\newcommand{\vtwo}{\mbox{$v_{2}$}\xspace}
\newcommand{\RP}{\mbox{$\Psi_{\rm{RP}}$}\xspace}
\newcommand{\NpOT}{\mbox{${N_{\rm part}^{1/3}}$}\xspace}
\newcommand{\KET}{\mbox{$KE_T$}\xspace}
\newcommand{\nq}{\mbox{$n_q$}\xspace}

\newcommand{\eps}{\mbox{${\varepsilon}$}\xspace}
\newcommand{\mT}{\mbox{$m_T$}\xspace}

\newcommand{\NcollOT}{\mbox{${N_{\rm coll}^{1/3}}$}\xspace}
\usepackage{graphicx}
\usepackage{xspace}
\usepackage{multirow}

\begin{document}
 
\title{Systematic Study of Azimuthal Anisotropy in Cu$+$Cu and Au$+$Au 
Collisions at $\sqrt{s_{_{NN}}}$=62.4 and 200~GeV}

\newcommand{\abilene}{Abilene Christian University, Abilene, Texas 79699, USA}
\newcommand{\augie}{Department of Physics, Augustana College, Sioux Falls, South Dakota 57197, USA}
\newcommand{\banaras}{Department of Physics, Banaras Hindu University, Varanasi 221005, India}
\newcommand{\baruch}{Baruch College, City University of New York, New York, New York, 10010 USA}
\newcommand{\bnlcoll}{Collider-Accelerator Department, Brookhaven National Laboratory, Upton, New York 11973-5000, USA}
\newcommand{\bnlphys}{Physics Department, Brookhaven National Laboratory, Upton, New York 11973-5000, USA}
\newcommand{\caucr}{University of California - Riverside, Riverside, California 92521, USA}
\newcommand{\charlesczech}{Charles University, Ovocn\'{y} trh 5, Praha 1, 116 36, Prague, Czech Republic}
\newcommand{\ciae}{Science and Technology on Nuclear Data Laboratory, China Institute of Atomic Energy, Beijing 102413, People's~Republic~of~China}
\newcommand{\cns}{Center for Nuclear Study, Graduate School of Science, University of Tokyo, 7-3-1 Hongo, Bunkyo, Tokyo 113-0033, Japan}
\newcommand{\colorado}{University of Colorado, Boulder, Colorado 80309, USA}
\newcommand{\columbia}{Columbia University, New York, New York 10027 and Nevis Laboratories, Irvington, New York 10533, USA}
\newcommand{\czechtech}{Czech Technical University, Zikova 4, 166 36 Prague 6, Czech Republic}
\newcommand{\dapnia}{Dapnia, CEA Saclay, F-91191, Gif-sur-Yvette, France}
\newcommand{\debrecen}{Debrecen University, H-4010 Debrecen, Egyetem t{\'e}r 1, Hungary}
\newcommand{\elte}{ELTE, E{\"o}tv{\"o}s Lor{\'a}nd University, H-1117 Budapest, P\'azmany P\'eter s\'et\'any 1/A, Hungary}
\newcommand{\fit}{Florida Institute of Technology, Melbourne, Florida 32901, USA}
\newcommand{\fsu}{Florida State University, Tallahassee, Florida 32306, USA}
\newcommand{\gsu}{Georgia State University, Atlanta, Georgia 30303, USA}
\newcommand{\hiroshima}{Hiroshima University, Kagamiyama, Higashi-Hiroshima 739-8526, Japan}
\newcommand{\ihepprot}{IHEP Protvino, State Research Center of Russian Federation, Institute for High Energy Physics, Protvino, 142281, Russia}
\newcommand{\illuiuc}{University of Illinois at Urbana-Champaign, Urbana, Illinois 61801, USA}
\newcommand{\inrras}{Institute for Nuclear Research of the Russian Academy of Sciences, prospekt 60-letiya Oktyabrya 7a, Moscow 117312, Russia}
\newcommand{\instpasczech}{Institute of Physics, Academy of Sciences of the Czech Republic, Na Slovance 2, 182 21 Prague 8, Czech Republic}
\newcommand{\isu}{Iowa State University, Ames, Iowa 50011, USA}
\newcommand{\jaea}{Advanced Science Research Center, Japan Atomic Energy Agency, 2-4 Shirakata Shirane, Tokai-mura, Naka-gun, Ibaraki-ken 319-1195, Japan}
\newcommand{\jinrdubna}{Joint Institute for Nuclear Research, 141980 Dubna, Moscow Region, Russia}
\newcommand{\kaeri}{KAERI, Cyclotron Application Laboratory, Seoul, Korea}
\newcommand{\kek}{KEK, High Energy Accelerator Research Organization, Tsukuba, Ibaraki 305-0801, Japan}
\newcommand{\korea}{Korea University, Seoul, 136-701, Korea}
\newcommand{\kurchatov}{Russian Research Center ``Kurchatov Institute," Moscow, 123098 Russia}
\newcommand{\kyoto}{Kyoto University, Kyoto 606-8502, Japan}
\newcommand{\labllr}{Laboratoire Leprince-Ringuet, Ecole Polytechnique, CNRS-IN2P3, Route de Saclay, F-91128, Palaiseau, France}
\newcommand{\lahorelums}{Physics Department, Lahore University of Management Sciences, Lahore 54792, Pakistan}
\newcommand{\lawllnl}{Lawrence Livermore National Laboratory, Livermore, California 94550, USA}
\newcommand{\losalamos}{Los Alamos National Laboratory, Los Alamos, New Mexico 87545, USA}
\newcommand{\lpc}{LPC, Universit{\'e} Blaise Pascal, CNRS-IN2P3, Clermont-Fd, 63177 Aubiere Cedex, France}
\newcommand{\lund}{Department of Physics, Lund University, Box 118, SE-221 00 Lund, Sweden}
\newcommand{\michigan}{Department of Physics, University of Michigan, Ann Arbor, Michigan 48109-1040, USA}
\newcommand{\muenster}{Institut f\"ur Kernphysik, University of Muenster, D-48149 Muenster, Germany}
\newcommand{\myongji}{Myongji University, Yongin, Kyonggido 449-728, Korea}
\newcommand{\nagasaki}{Nagasaki Institute of Applied Science, Nagasaki-shi, Nagasaki 851-0193, Japan}
\newcommand{\natmephi}{National Research Nuclear University, MEPhI, Moscow Engineering Physics Institute, Moscow, 115409, Russia}
\newcommand{\newmex}{University of New Mexico, Albuquerque, New Mexico 87131, USA }
\newcommand{\nmsu}{New Mexico State University, Las Cruces, New Mexico 88003, USA}
\newcommand{\ohio}{Department of Physics and Astronomy, Ohio University, Athens, Ohio 45701, USA}
\newcommand{\ornl}{Oak Ridge National Laboratory, Oak Ridge, Tennessee 37831, USA}
\newcommand{\orsay}{IPN-Orsay, Universite Paris Sud, CNRS-IN2P3, BP1, F-91406, Orsay, France}
\newcommand{\peking}{Peking University, Beijing 100871, People's~Republic~of~China}
\newcommand{\pnpi}{PNPI, Petersburg Nuclear Physics Institute, Gatchina, Leningrad Region, 188300, Russia}
\newcommand{\riken}{RIKEN Nishina Center for Accelerator-Based Science, Wako, Saitama 351-0198, Japan}
\newcommand{\rikjrbrc}{RIKEN BNL Research Center, Brookhaven National Laboratory, Upton, New York 11973-5000, USA}
\newcommand{\rikkyo}{Physics Department, Rikkyo University, 3-34-1 Nishi-Ikebukuro, Toshima, Tokyo 171-8501, Japan}
\newcommand{\saispbstu}{Saint Petersburg State Polytechnic University, St. Petersburg, 195251 Russia}
\newcommand{\saopaulo}{Universidade de S{\~a}o Paulo, Instituto de F\'{\i}sica, Caixa Postal 66318, S{\~a}o Paulo CEP05315-970, Brazil}
\newcommand{\seoulnat}{Department of Physics and Astronomy, Seoul National University, Seoul 151-742, Korea}
\newcommand{\stonybrkc}{Chemistry Department, Stony Brook University, SUNY, Stony Brook, New York 11794-3400, USA}
\newcommand{\stonycrkp}{Department of Physics and Astronomy, Stony Brook University, SUNY, Stony Brook, New York 11794-3800, USA}
\newcommand{\subatech}{SUBATECH (Ecole des Mines de Nantes, CNRS-IN2P3, Universit{\'e} de Nantes) BP 20722 - 44307, Nantes, France}
\newcommand{\tenn}{University of Tennessee, Knoxville, Tennessee 37996, USA}
\newcommand{\titech}{Department of Physics, Tokyo Institute of Technology, Oh-okayama, Meguro, Tokyo 152-8551, Japan}
\newcommand{\tsukuba}{Institute of Physics, University of Tsukuba, Tsukuba, Ibaraki 305, Japan}
\newcommand{\vandy}{Vanderbilt University, Nashville, Tennessee 37235, USA}
\newcommand{\waseda}{Waseda University, Advanced Research Institute for Science and Engineering, 17 Kikui-cho, Shinjuku-ku, Tokyo 162-0044, Japan}
\newcommand{\weizmann}{Weizmann Institute, Rehovot 76100, Israel}
\newcommand{\wigner}{Institute for Particle and Nuclear Physics, Wigner Research Centre for Physics, Hungarian Academy of Sciences (Wigner RCP, RMKI) H-1525 Budapest 114, POBox 49, Budapest, Hungary}
\newcommand{\yonsei}{Yonsei University, IPAP, Seoul 120-749, Korea}
\affiliation{\abilene}
\affiliation{\augie}
\affiliation{\banaras}
\affiliation{\baruch}
\affiliation{\bnlcoll}
\affiliation{\bnlphys}
\affiliation{\caucr}
\affiliation{\charlesczech}
\affiliation{\ciae}
\affiliation{\cns}
\affiliation{\colorado}
\affiliation{\columbia}
\affiliation{\czechtech}
\affiliation{\dapnia}
\affiliation{\debrecen}
\affiliation{\elte}
\affiliation{\fit}
\affiliation{\fsu}
\affiliation{\gsu}
\affiliation{\hiroshima}
\affiliation{\ihepprot}
\affiliation{\illuiuc}
\affiliation{\inrras}
\affiliation{\instpasczech}
\affiliation{\isu}
\affiliation{\jaea}
\affiliation{\jinrdubna}
\affiliation{\kaeri}
\affiliation{\kek}
\affiliation{\korea}
\affiliation{\kurchatov}
\affiliation{\kyoto}
\affiliation{\labllr}
\affiliation{\lahorelums}
\affiliation{\lawllnl}
\affiliation{\losalamos}
\affiliation{\lpc}
\affiliation{\lund}
\affiliation{\michigan}
\affiliation{\muenster}
\affiliation{\myongji}
\affiliation{\nagasaki}
\affiliation{\natmephi}
\affiliation{\newmex}
\affiliation{\nmsu}
\affiliation{\ohio}
\affiliation{\ornl}
\affiliation{\orsay}
\affiliation{\peking}
\affiliation{\pnpi}
\affiliation{\riken}
\affiliation{\rikjrbrc}
\affiliation{\rikkyo}
\affiliation{\saispbstu}
\affiliation{\saopaulo}
\affiliation{\seoulnat}
\affiliation{\stonybrkc}
\affiliation{\stonycrkp}
\affiliation{\subatech}
\affiliation{\tenn}
\affiliation{\titech}
\affiliation{\tsukuba}
\affiliation{\vandy}
\affiliation{\waseda}
\affiliation{\weizmann}
\affiliation{\wigner}
\affiliation{\yonsei}
\author{A.~Adare} \affiliation{\colorado}
\author{S.~Afanasiev} \affiliation{\jinrdubna}
\author{C.~Aidala} \affiliation{\columbia} \affiliation{\michigan}
\author{N.N.~Ajitanand} \affiliation{\stonybrkc}
\author{Y.~Akiba} \affiliation{\riken} \affiliation{\rikjrbrc}
\author{H.~Al-Bataineh} \affiliation{\nmsu}
\author{A.~Al-Jamel} \affiliation{\nmsu}
\author{J.~Alexander} \affiliation{\stonybrkc}
\author{K.~Aoki} \affiliation{\kyoto} \affiliation{\riken}
\author{L.~Aphecetche} \affiliation{\subatech}
% ...
\collaboration{PHENIX Collaboration} \noaffiliation


\begin{abstract}


We have studied the dependence of azimuthal anisotropy $v_2$ for 
inclusive and identified charged hadrons in Au$+$Au and Cu$+$Cu 
collisions on collision energy, species, and centrality. The values of 
$v_2$ as a function of transverse momentum $p_T$ and centrality in 
Au$+$Au collisions at $\sqrt{s_{_{NN}}}$=200~GeV and 62.4~GeV are the 
same within uncertainties. However, in Cu$+$Cu collisions we observe a 
decrease in $v_2$ values as the collision energy is reduced from 200 to 
62.4~GeV. The decrease is larger in the more peripheral collisions. By 
examining both Au$+$Au and Cu$+$Cu collisions we find that $v_2$ depends 
both on eccentricity and the number of participants, $N_{\rm part}$. We 
observe that $v_2$ divided by eccentricity ($\varepsilon$) monotonically 
increases with $N_{\rm part}$ and scales as ${N_{\rm part}^{1/3}}$. The 
Cu$+$Cu data at 62.4 GeV falls below the other scaled $v_{2}$ data. For 
identified hadrons, $v_2$ divided by the number of constituent quarks 
$n_q$ is independent of hadron species as a function of transverse 
kinetic energy $KE_T=m_T-m$ between $0.1<KE_T/n_q<1$~GeV. Combining all 
of the above scaling and normalizations, we observe a near-universal 
scaling, with the exception of the Cu$+$Cu data at 62.4 GeV, of 
$v_2/(n_q\cdot\varepsilon\cdot N^{1/3}_{\rm part})$ vs $KE_T/n_q$ for 
all measured particles.

\end{abstract}
 
 
\maketitle

\section{Introduction}

The azimuthal anisotropy of particles produced in relativistic heavy ion 
collisions is a powerful probe for investigating the characteristics of 
the quark-gluon plasma 
(QGP)~\cite{Adcox:2004mh,Adams:2005dq,Back:2004je,Arsene:2004fa}. 
The elliptic azimuthal anisotropy (\vtwo) is defined by 
the amplitude of the second-order harmonic in a Fourier series expansion 
of emitted particle azimuthal distributions:
\begin{equation}
\vtwo = \left<\cos{(2[\phi-\RP])}\right>,
\end{equation}
where $\phi$ represents the azimuthal emission angle of a particle and \RP 
is the azimuthal angle of the reaction plane, which is defined by the 
impact parameter and the beam axis. The brackets denote statistical 
averaging over particles and events. Elliptic flow is sensitive to the 
early stage of heavy ion collisions because pressure gradients transfer 
the initial geometrical anisotropy of the collision region to an 
anisotropy in momentum space.

One of the most remarkable findings at the Relativistic Heavy Ion Collider 
(RHIC) is that the strength of \vtwo~\cite{Adler:2003kt} is much 
larger than what is expected from a hadronic scenario 
\cite{Bleicher:2000sx}. Moreover, a scaling of \vtwo by the number of 
constituent quarks in a hadron in the intermediate transverse momentum 
region (\pt $ = 1$-$4$ GeV$/c$) has been found for a broad range of 
particle species produced in Au$+$Au at \sqsn=200~GeV 
\cite{Adare:2006ti,Adare:2012vq}. Both STAR and PHENIX 
experiments have observed that \vtwo scales better as a function of the 
transverse kinetic energy of the hadron. These scalings of \vtwo are 
consistent with constituent quark flow at early collision times and 
recombination as the dominant process of hadronization.

The detailed interpretation of \vtwo results requires modeling 
\cite{Heinz:2013th,Ollitrault:1992bk} of the wavefunction of the incoming 
nuclei, fluctuations of the initial geometry, viscous relativistic 
hydrodynamics, hadronic freeze out and subsequent rescattering, along with 
various model parameters such as the assumed equation of state and 
transport coefficients, e.g. viscosity. In recent calculations, the 
strength of \vtwo for hadrons in heavy ion collisions at \sqsn = 200 GeV 
can be reproduced by hydrodynamical models that include shear viscosity 
and initial 
fluctuations~\cite{Niemi:2012ry,Song:2011hk,Soltz:2012rk}.

At the LHC, experiments have measured \vtwo as a function of \pt from 
Pb$+$Pb collisions at an order of magnitude higher beam energy, at \sqsn = 
2.76~TeV~\cite{Aamodt:2010pb, ATLAS:2012at, Chatrchyan:2012ta}. These \vtwo results as a 
function of \pt for inclusive hadrons are very similar in magnitude and 
shape to the RHIC measurements at 200~GeV.  However, the \vtwo 
measurements for identified hadrons at 
LHC~\cite{Abelev:2012di,Abelev:2014pua} below 3 GeV/$c$ do not scale well 
with the quark number and transverse kinetic energy of the hadron with 
deviations up to 40\%.

A comparison of measured \vtwo at the lower beam energies at RHIC (\sqsn = 
7.7--200~GeV) shows that \vtwo as a function of \pt seems to be saturated 
above \sqsn $= 39$ GeV and decreases below this beam 
energy~\cite{Adamczyk:2013gw}. The scaling of \vtwo with transverse 
kinetic energy is broken below a beam energy of 
19~GeV~\cite{Adamczyk:2013gw}. 
Possible explanations for this behavior include
rescattering in the later hadronic phase, 
incomplete thermalization in the initial stage, or the plasma not being 
formed at these lower beam energies.

Because transverse kinetic energy scaling is broken at energies significantly 
lower and higher than RHIC's full energy of 200 GeV, it is important to provide 
systematic measurements of $v_2$ for identified hadrons as a function of system 
size, collision energy, and centrality. These systematics are needed in order 
to make progress on the nature of the QGP at lower energy-density. We report on 
such a set of measurements in this paper, examining both Au$+$Au and Cu$+$Cu 
collisions at 200~GeV and 62.4~GeV beam energies. This adds to the low-energy 
Au$+$Au measurements made by STAR~\cite{Adamczyk:2013gw} and their Cu$+$Cu 
\vtwo data at 200~GeV and 62.4~GeV beam energies~\cite{Abelev:2010tr}.  The 
system size dependence of flow is particularly important because long-range 
azimuthal correlations have also been observed in high-multiplicity events from 
much smaller systems such as $d$$+$Au collisions~\cite{Adare:2013piz} at RHIC, 
$p$$+$$p$~\cite{Khachatryan:2010gv}, and $p$$+$Pb collisions~\cite{CMS:2012qk} at LHC.  The 
origin of these anisotropies is currently unknown; various competing 
explanations include parton saturation and hydrodynamic flow.

We expect that the systematic study of \vtwo for inclusive and identified 
particles can provide information on the temperature dependence of 
$\eta/$s (\textit{i.e.} the ratio of shear viscosity to entropy density 
s), the impact of viscosity on systems of different sizes, as well as 
constraining models of the reaction dynamics.

The organization of this paper is as follows: 
Section II describes the PHENIX detector used for this analysis, 
Section III describes the experimental method of azimuthal anisotropy analysis, 
Section IV presents the results of the systematic study for inclusive 
charged hadron \vtwo, and 
Section V presents the results of the systematic study for the  $v_2$ of identified 
charged hadrons.
The new data published in this paper are the Cu$+$Cu data at 62.4~GeV, as well the Au$+$Au \vtwo results for \pt$>5~$GeV/$c$. 
Other data come from prior PHENIX publications.~\cite{Afanasiev:2009wq, Adare:2006ti}

\section{PHENIX DETECTOR}

The results that we present in this paper were obtained with the PHENIX 
detector at RHIC~\cite{Adcox:2003zm}. We discuss below the main detector 
components that were used for this analysis.

\begin{figure}[htbp]
\includegraphics[width=1.0\linewidth]{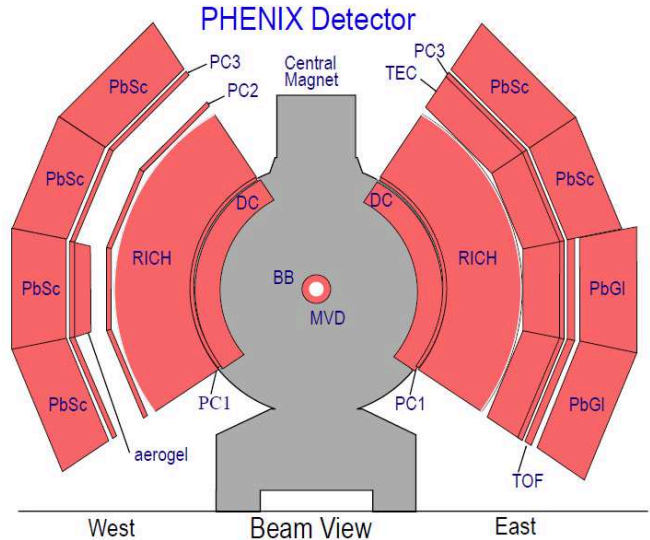}
\caption{\label{fig:in_detectors} (Color online)
Installed and active detectors for the RUN-4 configuration of the PHENIX 
experiment. Shown are the two central spectrometer arms viewed in a cut 
through the collision vertex.}
\end{figure}

\subsection{Global Detectors}

The Beam-Beam Counters (BBCs) are located 144 cm upstream and downstream 
of the beam crossing point.  Each BBC comprises 64 individual quartz 
\v{C}erenkov counters and covers the full azimuthal angle in the 
pseudorapidity range $3.0<|\eta|<3.9$. The average of the times measured 
by the two BBCs from fast leading particles provide the start time for the 
event, while the difference in times provides the vertex position of the 
collision. The timing and position resolution of the BBCs are 20 ps and 
0.6 cm respectively for both Au$+$Au and Cu$+$Cu collisions. The event 
start time is also used for particle identification through the 
time-of-flight to the TOF and EMCal subsystems in the PHENIX central arms.

The Zero Degree Calorimeters (ZDCs) cover the pseudorapidity range 
$|\eta|>6$ and measure the energy of spectator neutrons with an energy 
resolution of approximately 20\%. More details about these detectors can 
be found in \cite{Aizawa:2003zq}.

\subsection{Central-arm tracking detectors}

Two (identical) Drift Chambers (DC) are installed in the east and west 
arms of the PHENIX central detector and are located between 2.02 and 2.46 
m radial distance from the interaction point. Each of the two drift 
chambers extends 180 cm along the beam direction and subtends $\pi$/2 in 
azimuth. The momentum resolution for tracks reconstructed by the DC is 
0.7\%$\oplus$1.1\%$p$ (GeV/$c$). The position of the DCs relative to the 
other detectors in the central spectrometer is shown in 
Fig.~\ref{fig:in_detectors} and details of the DC construction and 
tracking performance can be found in~\cite{Adcox:2003en}.

The PHENIX pad chambers (PC) are multi-wire proportional chambers composed 
of three separate layers of pixel detectors.  Each pad chamber detector 
contains a single plane of wires in a gas volume bounded by two cathode 
planes.The innermost pad chamber plane, PC1, is located between the DC and 
a ring-imaging \v{C}erenkov counter (RICH) on both East and West arms, PC2 
is placed in back of the RICH on the West arm only, and PC3 is located in 
front of the Electromagnetic Calorimeters on both East and West arms.

The PC system determines space points outside the magnetic field and hence 
provides straight-line particle trajectories. They are the only 
nonprojective detectors in the central tracking system and thus are 
critical elements of the pattern recognition. PC1 is also essential for 
determining the three-dimensional momentum vector by providing the $z$ 
coordinate of each track at the exit of the DC. Details of the PC 
construction and their performance can be found in~\cite{Adcox:2003en}.

\subsection{Time-of-flight counters}

The PHENIX time-of-flight (TOF) detector serves as a particle 
identification device for charged hadrons. The time resolution for the 
BBC-TOF system is around 120 ps, which enables 2$\sigma$ separation of 
$\pi$/K up to 2.0 GeV/$c$. The length of the flight path of each track 
from the event vertex to the TOF detector is calculated by the momentum 
reconstruction algorithm. The length and time of flight are combined to 
identify the charged particles. The TOF is located between the PC3 and 
EMCal in the east arm and about 5.06 m away from the collision vertex. It 
covers $\mid$ $\eta$ $\mid$ $<$ 0.35 and azimuthal angle, $\Delta \phi$ = 
45$^{\circ}$. Details of the TOF construction and performance can be found 
in~\cite{Aizawa:2003zq}.

\subsection{Electromagnetic calorimeter}

The PHENIX EMCal was designed to measure the spatial position and energy 
of electrons and photons produced in heavy ion collisions.  The EMCal 
covers the full central spectrometer acceptance of $|\eta|<$ 0.35 and is 
installed in both arms, each subtending $90^{\circ}$ in azimuth, i.e. 
larger than the TOF acceptance. The EMCal comprises six sectors of 
lead-scintillator (PbSc) calorimeters and two sectors of lead-glass (PbGl) 
calorimeters. The PbGl is not used in this analysis, but we note that the 
TOF detector is in front of the PbGl so no PID coverage is lost. The PbSc 
is a sampling calorimeter and has a timing resolution of 400 ps for 
hadrons. The PbSc can be used to separate $\pi$/$K$ with 2$\sigma$ up to 
1.0 GeV/$c$. Details of the PbSc construction and performance are 
described in~\cite{Aphecetche:2003zr}.

\subsection{RICH}

A Ring Imaging \v{C}erenkov Counter (RICH) is installed on each of the 
PHENIX central arms.  Each RICH detector is a threshold gas \v{C}erenkov 
detector with a high angular segmentation filled with CO$_2$ gas. In this 
analysis we use the RICH to reject electrons by removing tracks that match 
to a RICH ring. It is noted that charged pions with \pt larger than 4 
GeV/$c$ also radiate in the CO$_2$ gas.


\section{experimental method}

\subsection{Data sets and event selection}

We measured Cu$+$Cu and Au$+$Au collisions at \sqsn = 62.4 and 200~GeV. 
The Cu$+$Cu data were taken during RHIC Run-5 (2005) and Au$+$Au data were 
taken during RHIC Run-4 (2004) running periods. We used a minimum bias 
trigger that was defined by a coincidence between the two BBCs and an 
energy threshold of one neutron in both ZDCs. The collision vertex along 
the beam direction, $z$, was measured by the BBC. The total number of 
minimum bias events that were analyzed after requiring an offline vertex 
cut of $|z|<30$~cm and selecting good runs are listed in 
Table~\ref{tab:dataset_event_selection}.

\begin{table}[tbh]
\caption{\label{tab:dataset_event_selection} 
Information on the data sets and event statistics.}
\begin{ruledtabular}
\begin{tabular}{cccc}
Year   & Species & Energy [GeV] & \# of events \\ \hline
2004  & Au$+$Au    & 200                  & $8.2\times 10^8$\\
2004  & Au$+$Au    & 62.4                 & $2.6\times 10^7$\\
2005  & Cu$+$Cu    & 200                  & $8.0\times 10^8$\\
2005  & Cu$+$Cu    & 62.4                 & $3.4\times 10^8$\\
\end{tabular}
\end{ruledtabular}
\end{table}

In Au$+$Au collisions at 200~GeV the centrality of the collision was 
determined by using the correlation of the total energy deposited in the 
ZDCs with the total charge deposited in the BBCs, as described 
in~\cite{Adler:2004zn}.  However, in 200~GeV Cu$+$Cu, 62.4~GeV 
Cu$+$Cu, and 62.4 GeV Au$+$Au collisions, the resolving power of the ZDCs 
is insufficient to significantly contribute to the centrality definition. 
Therefore, the total charge deposited in the BBCs is used to determine 
centrality in these collision systems, as described 
in~\cite{Adler:2004zn}. A Glauber model Monte-Carlo simulation of 
the each collision~\cite{Alver:2006wh, Miller:2007ri} was used to 
estimate the average number of participating nucleons \Np and participant 
eccentricity (\eps). This simulation includes modeling of the BBC and ZDC 
response. The eccentricty \eps is also known as the participant 
eccentricity and includes the effect of fluctuation from the initial 
participant geometry. Table~\ref{tab:npart_ecc_part_4systems} summarizes 
\Np, its systematic uncertainties ($\Delta$\Np), \eps and its systematic 
uncertainties ($\Delta$\eps).

\begin{table*}[ht]
\caption{\label{tab:npart_ecc_part_4systems}
Number of participants (\Np), its uncertainty ($\Delta$\Np), 
participant eccentricity (\eps) and its uncertainty ($\Delta$\eps) 
from Glauber Monte-Carlo calculations for 
Au$+$Au and Cu$+$Cu collisions at 200 and 62.4 GeV.}
\begin{ruledtabular}
\begin{tabular}{ccccccccccccccccc} 
centrality & 
\multicolumn{4}{c}{Au$+$Au 200 GeV} &
\multicolumn{4}{c}{Au$+$Au 62.4 GeV} & 
\multicolumn{4}{c}{Cu$+$Cu 200 GeV} &
\multicolumn{4}{c}{Cu$+$Cu 62.4 GeV} \\ 
bin    & \Np & $\Delta$\Np&  \eps & $\Delta$\eps[\%] &
         \Np & $\Delta$\Np  & \eps & $\Delta$\eps[\%] &
         \Np & $\Delta$\Np  & \eps & $\Delta$\eps[\%] &
         \Np & $\Delta$\Np  & \eps & $\Delta$\eps[\%] \\ 
\hline
 0\%--10\%   & 325.2 & 3.3 & 0.103 & 2.6  & 320.7 & 7.9 &  0.107 & 2.3 & 98.2 & 2.4 & 0.163 & 2.0 & 93.3 & 2.6 & 0.169 & 1.7 \\ 
10\%--20\%  & 234.6 & 4.7 & 0.200 & 2.5  & 230.7 & 9.2 & 0.207 & 2.2 & 73.6 & 2.5 & 0.241 & 3.0 & 71.1 & 2.4  & 0.248 & 2.6 \\ 
20\%--30\%  & 166.6 & 5.4 & 0.284 & 2.1  & 163.2 & 7.6 & 0.292 & 2.0 & 53.0 & 1.9 & 0.317 & 1.9 & 51.3 & 2.0 & 0.324 & 1.9 \\ 
30\%--40\%  &  114.2 & 4.4 & 0.356 & 1.7  & 113.0 & 5.6 & 0.365 & 1.8 & 37.3 & 1.6 & 0.401 & 1.9 & 36.2 & 1.8 & 0.408 & 1.6 \\ 
40\%--50\%  &  74.4    & 3.8 & 0.422 & 1.5  & 74.5  & 4.1 & 0.431 & 1.3 & 25.4 & 1.3 & 0.484 & 1.6 & 24.9 & 1.5  & 0.494 & 2.1 \\ 
50\%--60\%  &  45.5    & 3.3 & 0.491 & 1.1  & 45.9  & 3.1 & 0.498 & 1.0 & 16.7 & 0.9 & 0.579 & 1.4 & 16.1 & 0.9 & 0.587 & 1.5 \\ 
60\%--70\%  &  25.7    & 3.8 & 0.567 & 0.7  & 25.9  & 1.7 & 0.573 & 0.8 & 10.4 & 0.6 & 0.674 & 2.1 & &                 & 0.696 & 2.3 \\ 
70\%--80\%  &  13.4   & 3.0  &  0.666 & 1.2  & &                 & 0.678 & 1.1 & 6.4  & 0.5 & 0.721 & 1.7 & &                  & 0.742 & 1.6 \\ 
80\%--90\%  &    &                  & 0.726 & 2.8  & &                  & 0.740 & 2.2 & &               & 0.856 & 7.2 & &                  & 0.867 & 6.2 \\ 
\end{tabular}
\end{ruledtabular}
\end{table*}



   
\subsection{Track selection}

The analysis was performed for inclusive charged hadrons over the 
transverse momentum range $0.2 < p_{T} < 10$~GeV/$c$, and for identified 
charged particles (pions ($\pi^{+}+\pi^{-}$), kaons ($K^{+}+K^{-}$), and 
protons ($p+\bar{p}$)) in the momentum range up to $p_{T}$ 2.2, 3, and 
4~GeV/$c$ respectively.

The track reconstruction procedure is described in~\cite{Mitchell:2002wu}.  
Tracks reconstructed by the DC which do not originate from the event 
vertex have been investigated as background to the inclusive charged 
particle measurement. The main background sources include secondary 
particles from hadron decays and $e^{+}e^{-}$ pairs from the conversion of 
photons in the material between the vertex and the DC~\cite{Adler:2003au}. 
To minimize background originating from the magnets, reconstructed tracks 
are required to have a $z$-position less than $\pm 80$ cm when the tracks 
cross the outer radius of the DC. The DC is outside the central magnet 
field hence we can approximate reconstructed tracks through the 
central-arm detectors as straight lines. This enables tracks to be 
projected to outer detectors and matched to measured hits. Good tracks are 
required to be matched to a hit in the PC3, as well as in the EMCal, 
within 2.5~$\sigma$ of the expected hit location in both azimuthal and 
beam directions.
 
The Ring Imaging $\check{\rm C}$erenkov detector (RICH) also reduces the 
conversion background. For tracks with $p_{T} < $4~GeV/$c$ we apply a cut 
of $n_0<0$ where $n_0$ is the number of fired phototubes in the RICH ring. 
For $p_{T} > $4~GeV/$c$, we require tracks to have $E/p > $ 0.2, where $E$ 
denotes the energy deposited in the EMCal and $p_{T}$ is the transverse 
momentum of particles measured in the DC.  Because most of the background 
from photon conversion are low-momentum particles that were incorrectly 
reconstructed at higher momentum, when we require a large deposit of 
energy in the EMCal this suppresses the conversion background 
\cite{Adler:2005ad}.

\begin{figure}[htbp]
\includegraphics[width=0.998\linewidth]{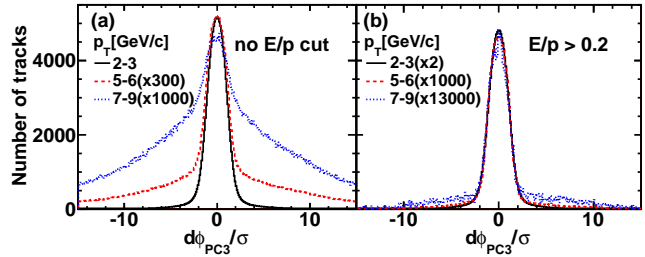}
\caption{\label{fig:h1_eovpcut} (Color online) 
(a) track/hit matching distribution of $d\phi/\sigma$ at PC3 without 
$E$/$p$ cut for indicated \pt bins; (b) same quantity, but after applying 
an $E/p > 0.2$ cut.}
\end{figure}

To demonstrate the effectiveness of the $E/p$ cut, Fig.~\ref{fig:h1_eovpcut} 
shows the track/hit matching distributions $d\phi/\sigma$ at PC3, where $d\phi$ 
is the residual between the track projection point and the detector hit 
position along $\phi$ and $\sigma$ is the standard deviation of the $d\phi$ 
distribution. The left panel shows the $d\phi/\sigma$ without an $E$/$p$ cut, 
and the right panel shows the distribution with a cut of $E/p>0.2$. Note that 
the vertical scale between the panels is different. The $E$/$p$ $>$ 0.2 cut 
substantially reduces the background for high \pt tracks.  The residual 
background remaining after these cuts has been estimated by the fitting the 
$d\phi/\sigma$ distributions in PC3 with a double Gaussian function (signal and 
background). The signal and residual background distributions are required to 
have the same mean. For $p_T<$~4~GeV/$c$ the residual background is less than 
5\% of the real tracks and reaches 10\% for $p_T$~8-10~GeV/$c$. The efficiency 
of the $E/p > 0.2$ cut is 0.3 at \pt $= 5 - 6$ GeV/$c$ and 0.1 at \pt $= 7 - 9$ 
GeV/$c$.

\subsection{Particle identification}

For identified charged hadrons we also require the tracks to have a hit in 
the TOF detector or EMCal within at most 2~$\sigma$ of the expected hit 
location in both azimuthal and beam directions. Particles are identified 
by their mass-squared, using the momentum measurement from the DC ($p$), 
time-of-flight between BBC and TOF/EMCal ($t$), and flight path length 
($L$) from the collision vertex point to the hit position on the TOF wall 
or cluster in the EMCal. The square of the particle's mass is calculated 
as
\begin{equation}
m^{2} = \frac{p^{2}}{c^{2}}\left[{\left(\frac{t}{L/c}\right)}^{2}-1\right]
\label{eq:m_square}
\end{equation}

\begin{figure}[htbp]
\includegraphics[width=1.0\linewidth]{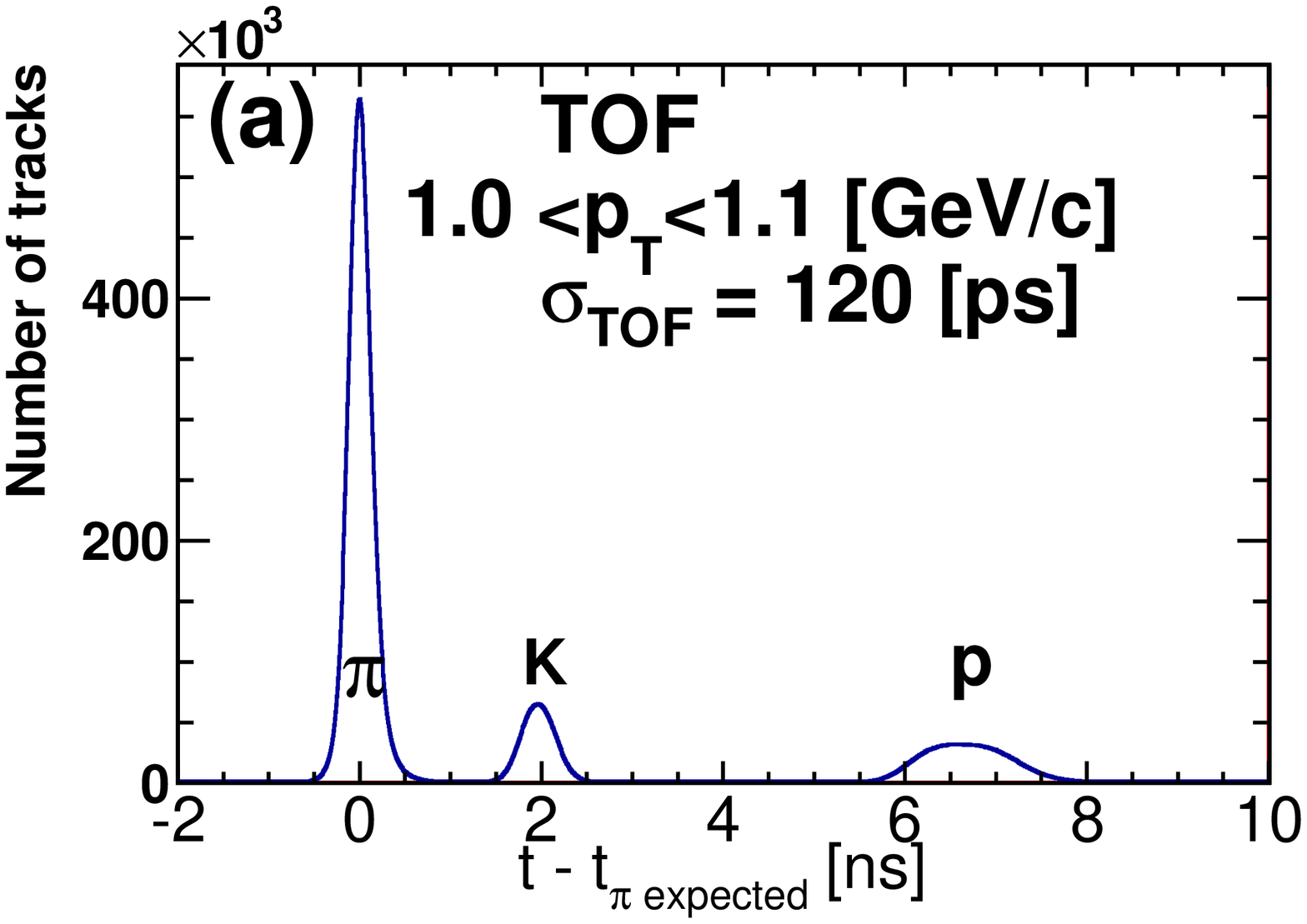}
\includegraphics[width=1.0\linewidth]{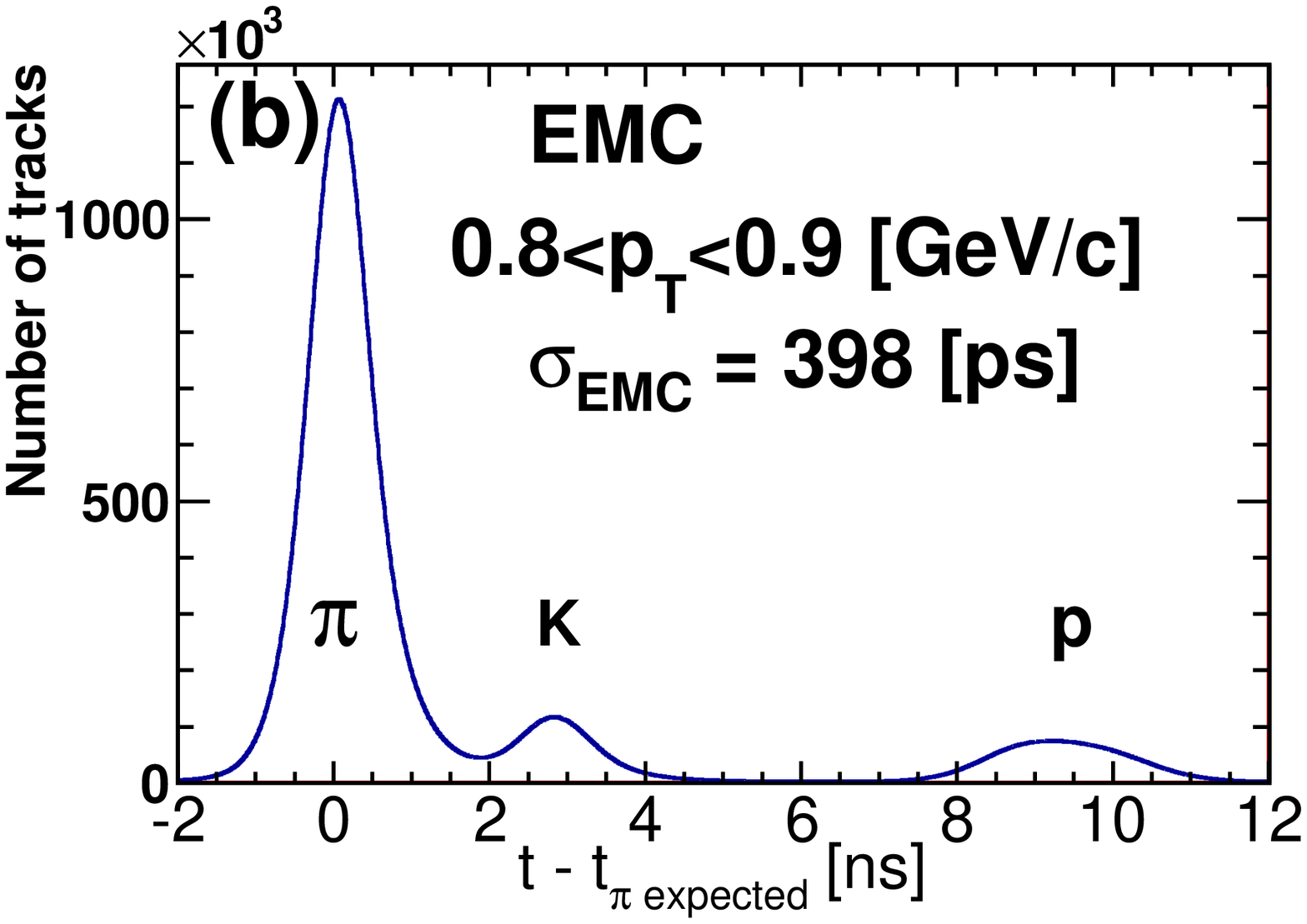}
\caption{\label{fig:T_Tpi_auau200_pid_tof_emc} (Color online)
Distributions of $t-t_{\pi {\rm expected}}$, the difference between the 
measured 
time-of-flight in the TOF (upper) and EMC (lower) and the time calculated 
assuming each candidate track is a pion. Resolutions are $\sigma_T \sim$ 
120~ps for TOF and $\sigma_T \sim$ 400~ps for EMCal in Au$+$Au at 200~GeV 
data. }
\end{figure}

\noindent The timing resolution of the BBC-TOF and BBC-EMCal systems was 
determined by examining the timing difference between the measured 
flight-time $t$ and $t_{\pi {\rm expected}}$, the time which is expected 
under the assumption that the particles are pions. The resulting time 
distribution is shown in Fig.~\ref{fig:T_Tpi_auau200_pid_tof_emc}.  A 
narrow peak centered around $t-t_{\pi {\rm expected}} \approx 0$ corresponds 
to pions, and the other two broad peaks are kaons and protons.  A Gaussian 
distribution is fit to the pion peak and yields a resolution of $\sim$ 120 
ps for the BBC-TOF system and $\sim$ 400 ps for the BBC-EMCal system.

The PID is performed by applying momentum-dependent cuts in mass-squared 
($m^2$). The $m^2$ distributions are fit with a 3-Gaussian function 
corresponding to pions, kaons, and protons.  The corresponding widths and 
centroids are extracted from the data as a function of transverse 
momentum. To select candidate tracks of a particle species, the $m^2$ is 
required to be within two standard deviations of the mean for the selected 
particles species and outside 2.5 standard deviations of the mean for the 
other particle species. This provides a sample for each particle species 
with at least 90\% purity in PID. For the BBC-TOF system the upper 
momentum cutoff is 2.2~GeV/$c$ for kaons and 3~GeV/$c$ for pions. For 
protons the upper momentum cutoff is 4~GeV/$c$. For the BBC-EMCal system 
the upper momentum cutoff is 1~GeV/$c$ for kaons and 1.4~GeV/$c$ for 
pions.  For protons the upper momentum cutoff is 2.2 GeV/$c$. The lower 
momentum cutoff for both PID systems is 0.2~GeV/$c$ for pions, 0.3~GeV/$c$ 
for kaons and 0.5~GeV/$c$ for protons. The PID results for the 200~GeV 
Au$+$Au data set were obtained using TOF detector only; for the 62.4~GeV 
Au$+$Au and 200~GeV Cu$+$Cu data sets the PID results were obtained by 
including identified particles from either the TOF or EMCal over different 
momentum ranges. For overlap region, we use BBC-EMC because of the better 
statistics and include the differences between BBC-EMC and BBC-TOF as 
systematic uncertainty shown in Tab.~\ref{tab:EMC_TOF_systematic_errors}. 
No correction is applied for any contamination from misidentified hadrons.

\subsection{Azimuthal anisotropy: event plane method \label{subsec:eventplanemethod}}

Because the principal axis of the participants cannot be measured directly 
in the experiment, the azimuthal angle of the reaction plane is 
estimated~\cite{Poskanzer:1998yz}. The estimated reaction plane is called 
the ``event plane'' and is determined for each harmonic of the Fourier 
expansion of the azimuthal distribution. The event flow vector $\vec{Q}_n 
= (Q_x, Q_y)$ and azimuth of the event plane $\Psi_n$ for $n$-th harmonic 
of the azimuthal anisotropy can be expressed as

\begin{eqnarray}  
Q_x & \equiv & |\vec{Q}_n| \cos{(n\Psi_n)} = \sum_i^M w_i \cos{(n\phi_i)}, \label{eq:flowvector_x} \\
Q_y & \equiv & |\vec{Q}_n| \sin{(n\Psi_n)} = \sum_i^M w_i \sin{(n\phi_i)}, \label{eq:flowvector_y} \\
\Psi_n & = & \frac{1}{n} \tan^{-1}\left(\frac{Q_y}{Q_x}\right), \label{eq:eventplane_definition}
\end{eqnarray}

\noindent
where $M$ denotes the number of particles used to determine the event 
plane, $\phi_i$ is the azimuthal angle of each particle and the
weight $w_i$ is the charge seen in the corresponding channel of the BBC.
Once the event plane is determined, the elliptic flow \vtwo can be 
extracted by correlating the azimuthal angle of emitted particles $\phi$ 
with the event plane:

\begin{eqnarray}
\vtwo \{ \Psi{n} \} = \frac{\vtwo^{\rm obs}}{{\rm Res}\{\Psi_{n}\}} 
= \frac{\left<\cos{(2[\phi-\Psi_n])}\right>}{\left<\cos{(2[\Psi_{n}-\RP])}\right>},
\label{eq:vtwo}
\end{eqnarray}

\noindent
where $\phi$ is the azimuthal angle of tracks in the laboratory frame, 
$\Psi_{n}$ is the $n$-th order event plane and the brackets denote an 
average over all charged tracks and events. The denominator 
Res\{$\Psi_n$\} is the event plane resolution that corrects for the 
difference between the estimated event plane $\Psi_n$ and true reaction 
plane \RP. We measure \vtwo using the same harmonic event plane 
($\Psi_{2}$) because this leads to a better accuracy~\cite{Poskanzer:1998yz}.

The second-harmonic event planes were independently determined with two 
BBCs located at forward (BBC South) and backward (BBC North) 
pseudorapidities $|\eta| = 3.1$--3.9~\cite{Adler:2003kt}. The 
planes were also combined to provide the event plane for the full event. 
More details study on using the BBC for the reaction plane measurement can 
be found in~\cite{Afanasiev:2009wq}. The measured \vtwo of hadrons in 
the central arms with respect to the combined second-harmonic BBC event 
plane will be denoted throughout this paper as \vtwo.


\subsubsection{Event plane determination}
\label{subsubsec:eventplane_determination}

To determine each event plane we chose the weights at each azimuthal angle 
to be the charge seen in the corresponding channel of the BBC. Corrections 
were performed to remove possible biases from small nonuniformities in the 
acceptance of the BBC. In this analysis we applied two corrections; the 
re-centering and shift methods~\cite{Poskanzer:1998yz}. In the 
re-centering method, event flow vectors are shifted and normalized using 
the mean $\left<Q\right>$ and width $\sigma$ of the $Q$ vector 
distribution;

\begin{eqnarray}
 Q_x' = \frac{Q_x-\left<Q_x\right>}{\sigma_x},\quad
 Q_y' = \frac{Q_y-\left<Q_y\right>}{\sigma_y}.
\end{eqnarray}

\noindent
This correction reduces the dependence of the event plane resolution on 
the laboratory angle. Most acceptance effects are removed by this 
re-centering method.  The shift method was used as a final 
correction~\cite{Poskanzer:1998yz}. In the shift method the reaction plane 
is shifted by $\Delta\Psi_n$ defined by

\begin{eqnarray}
n\Delta\Psi_n (\Psi_n) & = & \sum_{k=1}^{k_{\rm max}} \frac{2}{k} 
[-\left<\sin{(kn\Psi_n)}\right>\cos{(kn\Psi_n)} \nonumber \\
& & ~ + \left<\cos{(kn\Psi_n)}\right>\sin{(kn\Psi_n)}],
\label{eq:shift_correction}
\end{eqnarray}

\noindent
where $k_{\rm max}$ = 8 in this analysis. The shift ensures that 
$dN/d\Psi_n$ is isotropic. When $k_{\rm max}$ was reduced to 
$k_{\rm max}=4$, 
the difference in the extracted \vtwo was negligible and thus we 
include no systematic uncertainty due to the choice of $k_{\rm max}$ in 
our \vtwo results~\cite{Afanasiev:2009wq}.

Independent re-centering and shift corrections were applied to each 
centrality selection, in 5\% increments, as well as 20 cm steps in 
$z$-vertex. This optimizes the event plane resolution. The corrections 
were also performed for each experimental run (the duration of a run is 
typically 1-3 hours) to minimize the possible time-dependent response of 
detectors.

\subsubsection{Event plane resolution}
\label{subsubsec:eventplane_resolution}

The event plane resolution for \vtwo was evaluated by the two-subevent 
method. The event plane resolution~\cite{Poskanzer:1998yz} 
is expressed as

\begin{eqnarray}
& & \left<\cos{(kn[\Psi_n-\RP])}\right> 
= \frac{\sqrt{\pi}}{2\sqrt{2}}\chi_n e^{-\chi_n^2/4} \nonumber \\
& & \qquad \times \left[ I_{(k-1)/2}\left(\frac{\chi_n^2}{4}\right) + I_{(k+1)/2}\left(\frac{\chi_n^2}{4}\right) \right], 
\label{eq:resolution_formula}
\end{eqnarray}

\noindent
where $\chi_n = v_n\sqrt{2M}$, $M$ is the number of particles used to 
determine the event plane $\Psi_n$, $I_k$ is the modified Bessel function 
of the first kind and $k$ = 1 for the second harmonic BBC event plane.

To determine the event plane resolution we need to determine $\chi_n$. 
Because the North and South BBCs have approximately the same $\eta$ 
coverage, the event plane resolution of each sub-detector is expected to 
be the same. Thus, the subevent resolution for south and north event 
planes can be expressed as

\begin{eqnarray}
\left<\cos{(2[\Psi_n^{\rm S(N)}-\RP])}\right>
 = \sqrt{ \left<\cos{(2[\Psi_n^{\rm S}-\Psi_n^{\rm N}])}\right> },
 \label{eq:eventplane_resolution_twosubevents}
\end{eqnarray}

\noindent
where $\Psi_n^{\rm S(N)}$ denotes the event plane determined by the South 
(North) BBC. Once the subevent resolution is obtained from 
Eq.~(\ref{eq:eventplane_resolution_twosubevents}), one can calculate 
$\chi_n^{\rm sub}$ using Eq.~(\ref{eq:resolution_formula}). The $\chi_n$ 
for the full event can then be estimated by $\chi_n = \sqrt{2}\chi_n^{\rm 
sub}$. This is then substituted into Eq.~(\ref{eq:resolution_formula}) to 
give the full event resolution. Because the multiplicity of the full event 
is twice as large as that of the subevent, $\chi_n$ is proportional to 
$\sqrt{M}$.


\begin{figure}[htbp]
\includegraphics[width=1.0\linewidth]{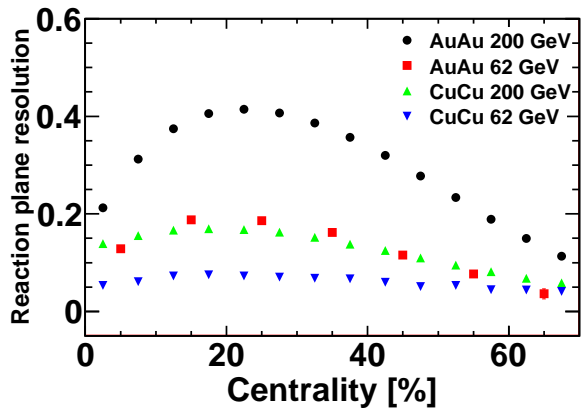}
\caption{\label{fig:RP_reso} (Color online)
Second-order event plane resolution vs. centrality in Au$+$Au and Cu$+$Cu 
at 200 and 62.4 GeV. The event plane is measured by BBC.}
\end{figure}

Figure~\ref{fig:RP_reso} shows the BBC North-South-combined resolution of 
the event plane as a function of the centrality in Au$+$Au and Cu$+$Cu at 
\sqsn ~= 200 and 62.4 GeV. The reaction-plane resolution and its 
uncertainties in Au$+$Au and Cu$+$Cu at 62.4 and 200~GeV are summarized in 
Table~\ref{tab:RP_resolution}.

\begingroup \squeezetable
\begin{table*}[tbh]
\caption{\label{tab:RP_resolution} 
Reaction-plane resolution for each centrality in Au$+$Au and Cu$+$Cu 
collisions at \sqsn~=~200 and 62.4 GeV and its statistical contribution to 
the uncertainty on \vtwo.  Note: Centrality bins are 10\% wide (0\%--10\%, 
10\%--20\%, etc.) for Au$+$Au 62.4 GeV.}
\begin{ruledtabular} \begin{tabular}{ccccccccc}
 &
\multicolumn{2}{c}{Au$+$Au 200 GeV} &
\multicolumn{2}{c}{Au$+$Au 62.4 GeV} &
\multicolumn{2}{c}{Cu$+$Cu 200 GeV} &
\multicolumn{2}{c}{Cu$+$Cu 62.4 GeV} \\
Centrality  & 
Reso- & Stat. Uncert. &
Reso- & Stat. Uncert. &
Reso- & Stat. Uncert. &
Reso- & Stat. Uncert. \\
bin & 
lution & for \vtwo [\%] &
lution & for \vtwo [\%] &
lution & for \vtwo [\%] &
lution & for \vtwo [\%] \\
\hline
   0\%--5\% & 0.212 & 0.20  & 0.128 &  2.0 & 0.139 & 0.55 & 0.053 & 5.6  \\
  5\%--10\% & 0.312 & 0.09  &       &        & 0.155 & 0.44 & 0.061 & 4.3  \\
 10\%--15\% & 0.375 & 0.06  & 0.189 &  0.94 & 0.167 & 0.38 & 0.073 & 3.0  \\
 15\%--20\% & 0.405 & 0.05  &       &        & 0.170 & 0.37 & 0.075 & 2.8  \\
 20\%--25\% & 0.414 & 0.05  & 0.186 &  0.97 & 0.168 & 0.38 & 0.073 & 3.0  \\
 25\%--30\% & 0.407 & 0.05  &       &        & 0.162 & 0.40 & 0.071 & 3.2  \\
 30\%--35\% & 0.387 & 0.06  & 0.163 &  1.3 & 0.152 & 0.46 & 0.068 & 3.4  \\
 35\%--40\% & 0.357 & 0.07  &       &        & 0.138 & 0.56 & 0.067 & 3.5  \\
 40\%--45\% & 0.320 & 0.09  & 0.118 &  2.4 & 0.125 & 0.68 & 0.060 & 4.4  \\
 45\%--50\% & 0.278 & 0.12  &       &        & 0.110 & 0.88 & 0.051 & 6.1  \\
 50\%--5\%5 & 0.234 & 0.16  & 0.079 &  5.4 & 0.095 & 1.2 & 0.054 & 5.6  \\
 55\%--60\% & 0.189 & 0.25  &       &        & 0.082 & 1.6 & 0.045 & 7.9  \\
 60\%--65\% & 0.150 & 0.40  & 0.044 & 17.5 & 0.068 & 2.3 & 0.044 & 8.2  \\
 65\%--70\% & 0.113 & 0.70  &       &        & 0.058 & 3.1 & 0.041 & 9.6  \\
\end{tabular} \end{ruledtabular}
\end{table*}
\endgroup

\subsection{Systematic uncertainty for \vtwo}

The sources of systematic uncertainty on the \vtwo measurement
include: reaction plane determination, the effects of matching cuts,
the effects of the E/p cut, and occupancy effects for PID \vtwo. These are described below.

\begin{table}[tbh]
\caption{\label{tab:RP_systematic_errors}
Systematic uncertainty [\%] of the reaction plane determination for each data 
set and each centrality bin. These are obtained by taking the larger 
values away from unity of the ratio of \vtwo with BBC North and South to 
\vtwo with BBC North-South-combined.
}
\begin{ruledtabular} \begin{tabular}{ccccc}
Centrality  &
\multicolumn{2}{c}{Au$+$Au} & 
\multicolumn{2}{c}{Cu$+$Cu} \\
bin & 
200 GeV & 
62.4 GeV & 
200 GeV & 
64 GeV \\
\hline 
 0\%--10\%   & 2 & 3 & 3 & 14  \\
10\%--20\%  & 3 & 2 & 2 & 9   \\
20\%--30\%  & 4 & 2 & 2 & 6   \\
30\%--40\%  & 4 & 7 & 2 & 2   \\
40\%--50\%  & 3 & 7 & 2 & 3   \\
50\%--60\%  & 3 & 5 & 2 & 5   \\
\end{tabular} \end{ruledtabular}
\end{table}

\begin{figure}[htbp]
\includegraphics[width=1.0\linewidth]{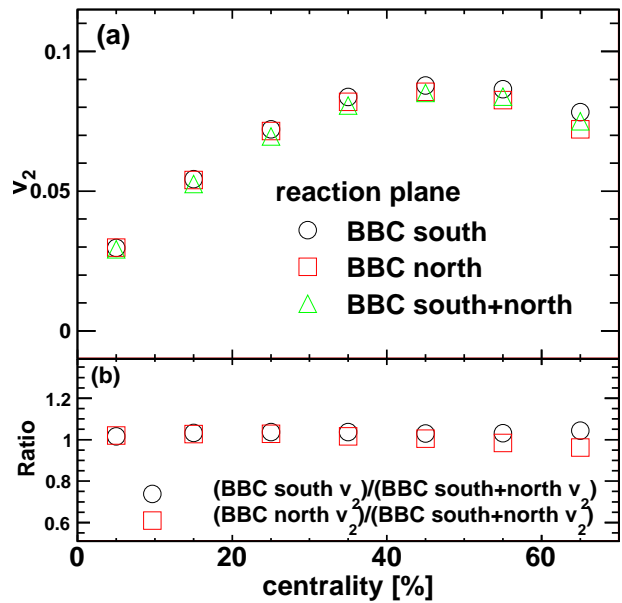}
\caption{\label{fig:RP_diff_Au200}  (Color online)
(a) \vtwo vs. centrality with three different reaction planes (BBC South, 
North, South-North combined) for Au$+$Au 200~GeV.  (b) The ratio of 
\vtwo with BBC South or North reaction plane to \vtwo with South-North 
combined. }
\end{figure}

The systematic uncertainties due to the reaction plane determination were 
estimated by comparing the \vtwo values extracted using three different 
reaction planes; the BBC North, BBC South, and BBC North-South combined. 
Figure~\ref{fig:RP_diff_Au200}a shows \vtwo ~vs. centrality for three 
reaction planes (BBC South, North, South-North combined) for 
Au$+$Au~200~GeV. The bottom panel shows the ratio of \vtwo with BBC North 
and South RP to \vtwo with BBC North-South combined (default). The 
percentage systematic uncertainty was obtained by taking the largest 
values away from unity of these ratios. These uncertainties are summarized 
in Table~\ref{tab:RP_systematic_errors} summarizes for each data set and 
each centrality bin.

\begingroup \squeezetable
\begin{table*}[tbh]
\caption{\label{tab:systematic_errors_MB}
Systematic uncertainty [\%] of the matching and $E/p$ cuts for each data 
set and each \pt~bin for minimum bias event sample, which are obtained by 
taking the larger values of the ratio of \vtwo with different matching cut 
to \vtwo with the default matching cut.}
\begin{ruledtabular} \begin{tabular}{ccccccccccccc}
      & &
\multicolumn{2}{c}{Au$+$Au 200 GeV} & &
\multicolumn{2}{c}{Au$+$Au 62.4 GeV} & &
\multicolumn{2}{c}{Cu$+$Cu 200 GeV} & &
\multicolumn{2}{c}{Cu$+$Cu 62.4 GeV} \\
 \pt  & & 
\multicolumn{2}{c}{Systematic Uncertainty (\%) } & &
\multicolumn{2}{c}{Systematic Uncertainty (\%) } & & 
\multicolumn{2}{c}{Systematic Uncertainty (\%) } & & 
\multicolumn{2}{c}{Systematic Uncertainty (\%) } \\ 
 (GeV/$c$) & &
Matching cut & $E/p$ cut & & 
Matching cut & $E/p$ cut & &
Matching cut & $E/p$ cut & &
Matching cut & $E/p$ cut \\
\hline
 0.2--1.0 & & 1 & 1 & & 1 & 2 & & 1 & 3 & & 2 & 3 \\  
 1.0--2.0 & & 1 & 3 & & 1 & 4 & & 1 & 2 & & 1 & 2 \\
 2.0--4.0 & & 1 & 2 & & 4 & 3 & & 1 & 3 & & 2 & 3 \\ 
\end{tabular} \end{ruledtabular}
\end{table*}
\endgroup


The default matching cuts for tracks projected to PC3 are $-2.5\sigma < 
(d\phi_{PC3}$ and $dz_{PC3}) < 2.5\sigma$. To obtain the systematic 
uncertainty from the dependence on these matching cuts, we examined 
different cut windows, e.g.  $|d\phi_{PC3}|< 1.0\sigma $ and $1.0\sigma < 
|d\phi_{PC3}| < 2.5\sigma$, and compared \vtwo values using these cuts to 
\vtwo values from the default cut. The difference between \vtwo values 
with these matching cuts determine the systematic uncertainties. Because 
the alternative cut windows have a smaller sample of data, we extracted 
the systematic uncertainty from the minimum bias event sample and used 
these for all centralities. Table~\ref{tab:systematic_errors_MB} shows the 
matching systematic uncertainties.


The $E/p$ cut can reject background from conversions, especially for high 
\pt tracks.  The default cut, $E/p > 0.2$, was used for tracks with 
\pt~$>$~4~GeV/$c$. To test the sensitivity to the value of the cut, we 
apply cuts of $E/p >$~0.1, 0.2 and 0.3 cuts for tracks 
$3<$\pt~$<$~4~GeV/$c$; a lower momentum was used because we have more 
statistics there. The ratio of \vtwo with different $E/p$ cuts contributes 
to the systematic uncertainty. We obtained the systematic uncertainty due 
to the $E/p$ cut using the minimum bias event sample, because within the 
statistics we did not observe any centrality dependence for how \vtwo 
changed with different $E/p$ cuts. Table~\ref{tab:systematic_errors_MB} 
lists the systematic uncertainties from the $E/p$ cut.


Both EMCal and TOF detectors are used for particle identification. In the 
low \pt region both detectors can be used, and the difference between 
\vtwo measured with the EMCal and TOF, averaged across \pt, is used for 
the systematic uncertainty due to timing performance. This includes the 
1\% uncertainty due to background contributions in the particle 
identification. The values are summarized in 
Table~\ref{tab:EMC_TOF_systematic_errors}.  Note, that the timing 
systematic uncertainty only affects the identified hadron results.

\begin{table}[tbh]
\caption{\label{tab:EMC_TOF_systematic_errors}
Systematic uncertainty [\%] for $v_2$ of identified hadrons due to the 
timing performance of the EMCal and TOF detectors. These are obtained by 
taking the difference between \vtwo with EMCal and \vtwo with TOF merging 
\pt and centrality bins.
}
\vspace{0.2cm}
\begin{ruledtabular}
\begin{tabular}{ccccc}
Collision  & \sqsn & \multicolumn{3}{c}{identified hadron} \\ 
Species    & [GeV] &  $\pi$ & K & p \\ \hline
Au$+$Au & 62.4 & 2 & 4 & 6 \\
Cu$+$Cu & 200  & 3 & 5 & 6 \\
\end{tabular}
\end{ruledtabular}
\end{table}
   
The values for \vtwo can be impacted due to finite occupancy which tends 
to lower the measured \vtwo. The magnitude of this effect has been 
estimated to be largest for central Au$+$Au collisions at 200~GeV as a 
reduction in \vtwo for PID particles of approximately 0.0013 for the 
running conditions of the data presented here. This effect is independent 
of \pt. For different centrality and beam-energies we take the systematic 
uncertainty on PID \vtwo to linearly decrease with the average charged 
particle multiplicity in those collisions.

\section{results for {\vtwo} of inclusive charged hadrons}

In this section we describe the \vtwo measurements and how they change as 
a function of collision energy and system size. We present the 
measured \vtwo for inclusive charged particles in Au$+$Au and Cu$+$Cu 
collisions at 62.4 and 200~GeV. For 200~GeV, the \vtwo results for 
\pt$<5~$GeV/$c$ are obtained by re-binning the data published 
in~\cite{Adare:2006ti,Afanasiev:2009wq,Adare:2010ux}. 
The new 200~GeV data published in this paper are 
\vtwo results for \pt$>5~$GeV/$c$. In addition the 62.4~GeV Cu$+$Cu data 
are new results original in this paper.

The centrality selections of each collision system are:

\begin{enumerate} 
 \item Au$+$Au collisions at \sqsn = 200 GeV  
   \begin{itemize}
     \item Minimum Bias ; 0\%--92\% 
     \item 10\% steps ; 0\%--10\%, 10\%--20\%, 20\%--30\%, 30\%--40\%, 
40\%--50\%, 50\%--60\%  
     \item 20\% steps ; 0\%--20\%, 20\%--40\%, 40\%--60\%  
     \item Most peripheral bin ; 60\%--92\% 
   \end{itemize}
 \item Au$+$Au collisions at \sqsn = 62.4 GeV  
 \begin{itemize}
     \item Minimum Bias ; 0\%--83\% 
     \item 10\% steps ; 0\%--10\%, 10\%--20\%, 20\%--30\%, 30\%--40\%, 40\%--50\%  
 \end{itemize}
 \item Cu$+$Cu collisions at \sqsn = 62.4  and 200 GeV  
 \begin{itemize}
     \item Minimum Bias ; 0\%--88\% 
     \item 10\% steps ; 0\%--10\%, 10\%--20\%, 20\%--30\%, 30\%--40\%, 
40\%--50\%
 \end{itemize}
\end{enumerate}    

\noindent

\subsection{\vtwo vs. \pt results for inclusive charged hadrons}
   
\subsubsection{Au$+$Au at \sqsn = 200GeV}

\begin{figure}[htbp]
\includegraphics[width=1.0\linewidth]{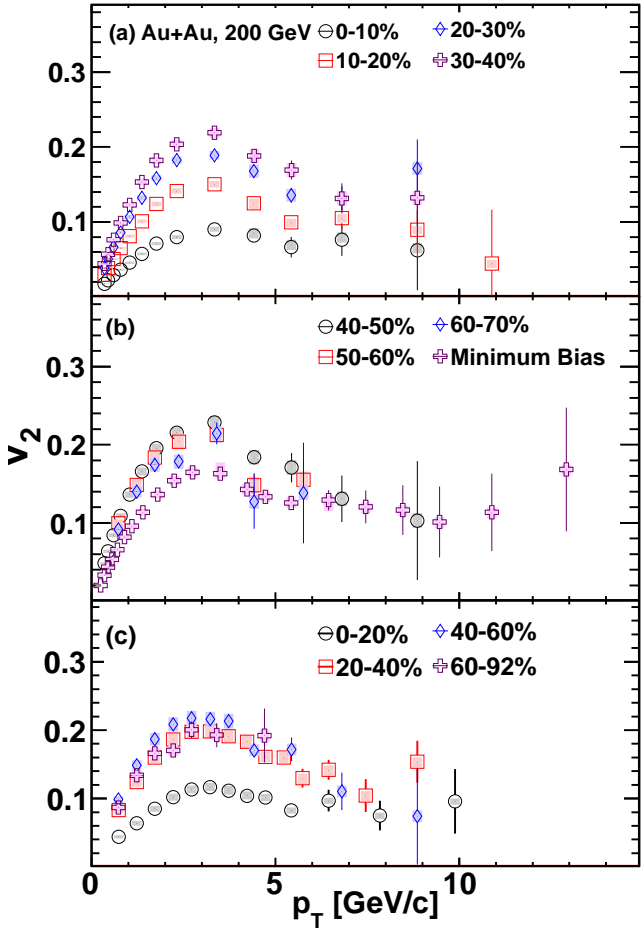}
\caption{\label{fig:inclusive_auau200}  (Color online)
\vtwo for inclusive charged hadrons in Au$+$Au at \sqsn=200~GeV for the 
centralities indicated. The error bars show statistical uncertainties and 
the bands show systematic uncertainties. In many cases, the systematic uncertainties are smaller than the symbols.}
\end{figure}

We analyzed 860 million Au$+$Au collisions at 200~GeV collected during the 
2003-04 experimental period, which is more than 20 times larger than the 
sample of events (30~M) analyzed from the 2001-02 experimental 
period~\cite{Adler:2003kt}. Figure~\ref{fig:inclusive_auau200} 
shows the \vtwo for inclusive charged hadrons in Au$+$Au collisions at 
200~GeV.

\subsubsection{Au$+$Au at \sqsn = 62.4GeV}

\begin{figure}[htbp]
\includegraphics[width=1.0\linewidth]{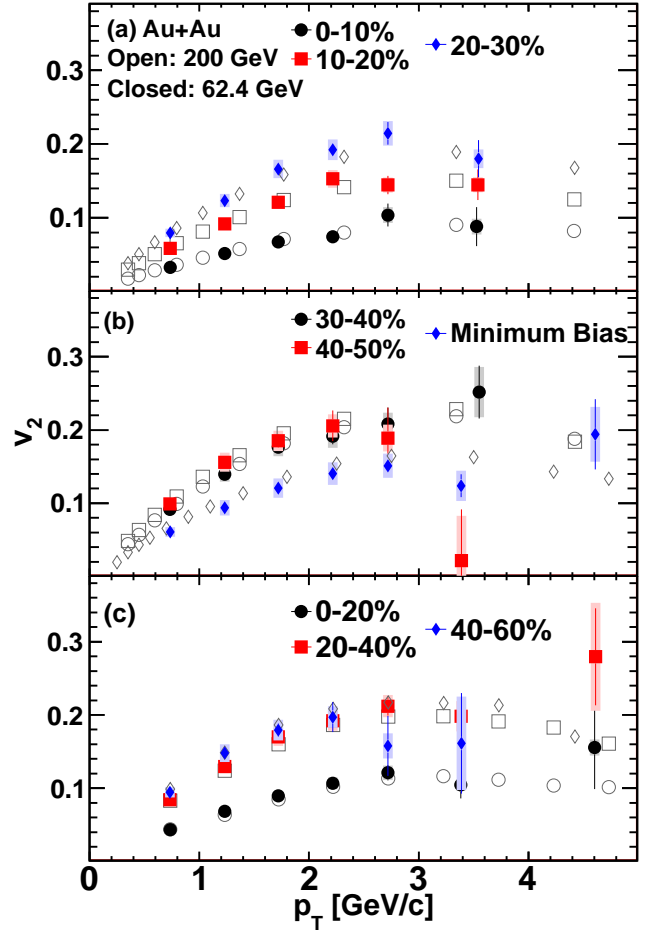}
\caption{\label{fig:inclusive_auau62} (Color online)
\vtwo for inclusive charged hadrons in Au$+$Au at \sqsn = 62.4 and 200 GeV 
for the centralities indicated. The error bars show statistical 
uncertainties and the bands show systematic uncertainties. In many cases, the systematic uncertainties are smaller than the symbols.}
\end{figure}

For Au$+$Au collisions at 62.4 GeV, 30 million events were analyzed to 
study the dependence of \vtwo on collision center-of-mass energy. The 
measured \vtwo results from this collision system are shown in 
Fig.~\ref{fig:inclusive_auau62}, together with the results from Au$+$Au 
200~GeV collisions. The values of \Np are very similar at these two beam 
energies. We observe that the \vtwo measurements for Au$+$Au collisions at 
62.4 GeV are consistent with those for Au$+$Au at 200 GeV, within the 
combined statistical and systematic uncertainties.

\subsubsection{Cu$+$Cu at \sqsn = 200 and 62.4GeV}


\begin{figure*}[htbp]
\includegraphics[width=0.998\linewidth]{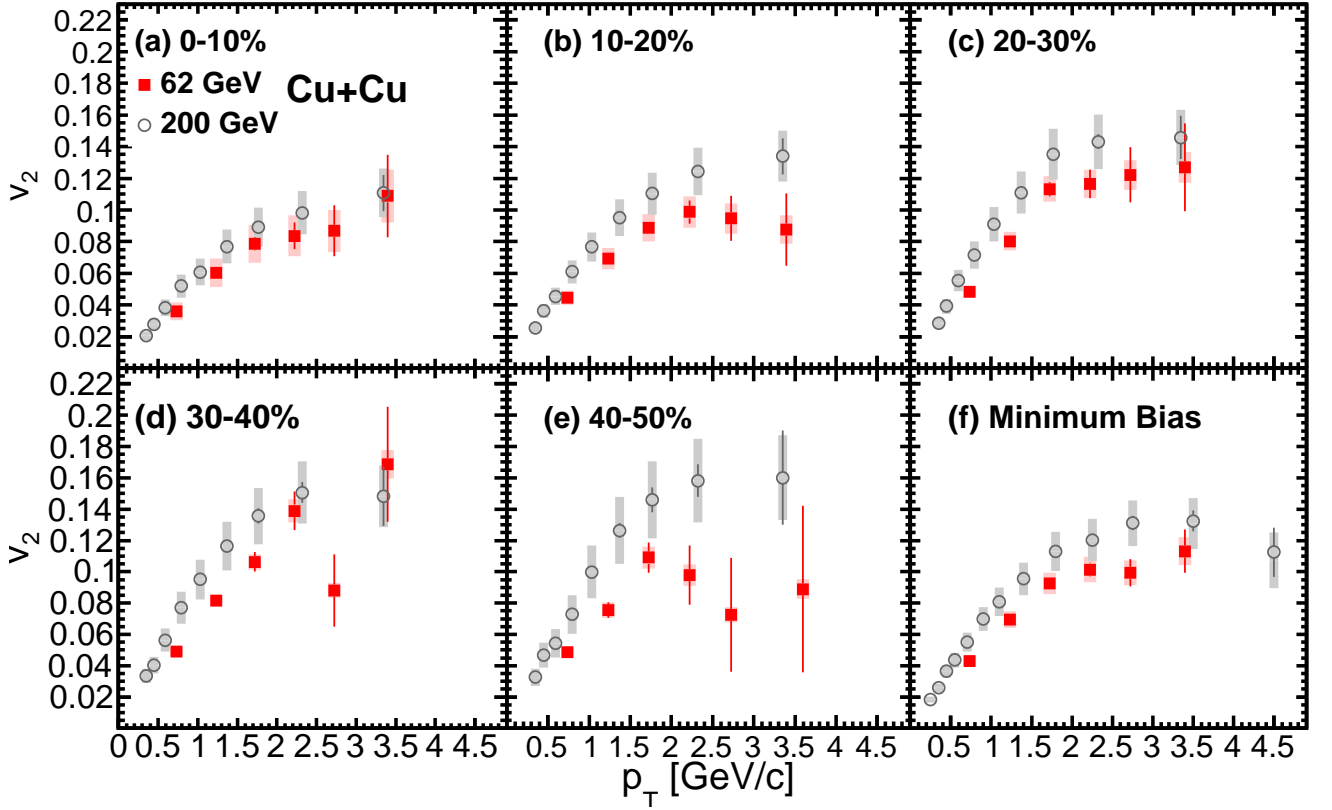}
\caption{\label{fig:inclusive_CuCu62} (Color online)
\vtwo for inclusive charged hadrons in Cu$+$Cu at \sqsn = 62.4 GeV 
compared 
with 200~GeV~\cite{Adare:2006ti} for the centralities indicated. 
The error bars show statistical uncertainties and the bands show 
systematic uncertainties. In many cases, the systematic uncertainties are smaller than the symbols.}
\end{figure*}

For Cu$+$Cu collisions at 62.4~GeV, 340 million events were analyzed to 
study the dependence of \vtwo on collision center-of-mass energy and 
system size. Figure~\ref{fig:inclusive_CuCu62} shows the \vtwo results at 
62.4~GeV in minimum bias events and 10\% centrality selections. These are 
compared with Cu$+$Cu~200~GeV \vtwo results~\cite{Adare:2006ti}. 
The \vtwo results for Cu$+$Cu collisions at 62.4~GeV are clearly smaller 
than those in 200~GeV collisions, especially at $\pt<1.5$~GeV/$c$.

     \subsection{System comparisons}
     \label{subsection:SystemComparisons}

\subsubsection{Centrality and collision energy dependence}

\begin{figure}[htbp]
\includegraphics[width=1.0\linewidth]{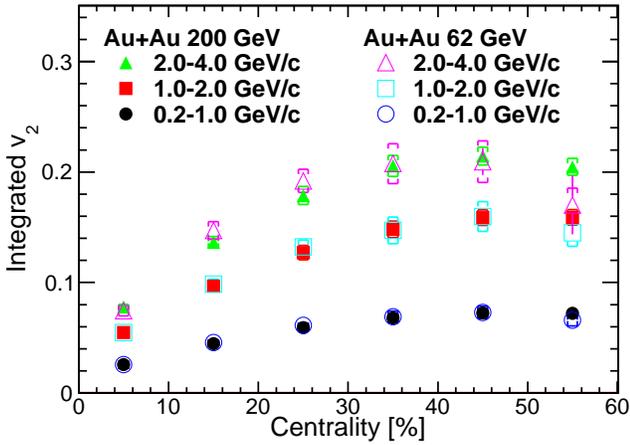}
\caption{\label{fig:v2_AuAu_cent_h} (Color online)
Comparison of integrated \vtwo at \sqsn = 62.4 and 200~GeV in Au$+$Au. 
Solid symbols indicate \sqsn = 200 GeV and open symbols indicate \sqsn = 
62.4 GeV.  Ranges of \pt integrated are 0.2--1.0 (circles), 1.0--2.0 
(squares), and 2.0--4.0 (triangles) GeV/$c$. The bars indicate the 
statistical uncertainties and the boxes indicate the systematic 
uncertainties. In many cases, the systematic uncertainties are smaller than the symbols.}
\end{figure}

\begin{figure}[htbp]
\includegraphics[width=1.0\linewidth]{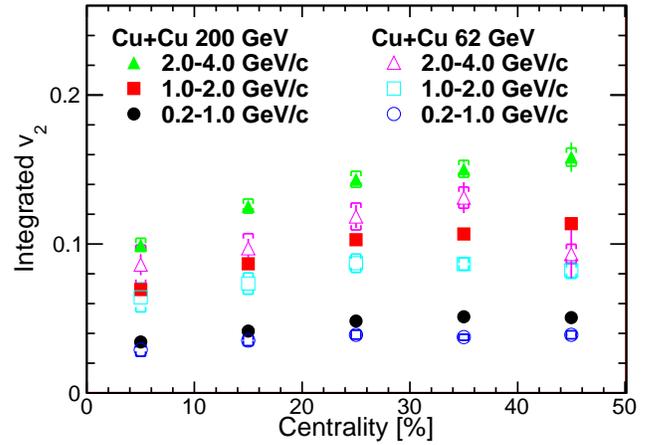}
\caption{\label{fig:v2_CuCu_cent_h} (Color online)
Comparison of integrated \vtwo at \sqsn = 62.4 and 200~GeV in Cu$+$Cu. Open 
symbols indicate \sqsn = 62.4 GeV and filled symbols indicate \sqsn = 200 
GeV.  Ranges of \pt integrated are 0.2--1.0 (circles), 1.0--2.0 
(squares), and 2.0--4.0 (triangles) GeV/$c$. The bars indicate the 
statistical uncertainties and the boxes indicate the systematic 
uncertainties. In many cases, the systematic uncertainties are smaller than the symbols.}
\end{figure}

An alternative view of these data is to make separate \pt selections and 
to plot \vtwo in a given \pt range as a function of centrality and 
collision energy. Figure~\ref{fig:v2_AuAu_cent_h} presents the Au$+$Au 
data as a function of centrality, where triangles, boxes, and circles 
correspond to three \pt bins: 0.2--1.0, 1.0--2.0 and 2.0--4.0~GeV/$c$ 
respectively. The two different beam energies are presented by open and 
closed symbols for 62.4 and 200~GeV respectively.  The data confirms prior 
results that \vtwo increases from central to midcentral collisions and 
then begins to decrease again towards peripheral collisions. The \vtwo for 
Au$+$Au at 62.4 and 200~GeV agree to within statistical and systematic 
uncertainties for all measured centralities.

A similar \vtwo comparison has been carried out by the STAR experiment 
reaching even lower energies from \sqsn $=$ 7.7 to 
200~GeV~\cite{Adamczyk:2013gw}. Their results show that the \vtwo(\pt) 
increases slightly from 7.7 up to 39~GeV, then saturates above 39~GeV.

Figure~\ref{fig:v2_CuCu_cent_h} shows the centrality dependence of \vtwo 
for charged hadrons emitted at different \pt from Cu$+$Cu collisions at 
62.4 and 200~GeV. The statistical uncertainties are larger due to lower 
statistics for the Cu$+$Cu in the 62.4~GeV data sample.
The measured \vtwo values are lower at 62.4~GeV compared with 200~GeV.

We have made a comparison between the measured PHENIX \vtwo and the previously 
published STAR \vtwo measurement~\cite{Abelev:2010tr}  in Cu+Cu collisions
and found them to be generally consistent.
 For 200~GeV Cu+Cu the PHENIX \vtwo are higher by about 10\% in
 the 0-10\%, 10-20\%, 20-30\% and 30-40\% centrality bins, and higher by 
about 20\% in 40-50\% bin; these differences are within statistical and 
systematic uncertainties  of the PHENIX results in all cases.
At 62.4~GeV the PHENIX \vtwo is lower by approximately
10\% in the 0-40\% bins and by 20\% in 40-50\% bin.
These differences are within statistical and systematic uncertainties in 
the 0-20\% bins, though they are roughly twice the statistical
and systematic uncertainties in 20-50\% bins, taking into account
errors on the PHENIX measurement alone.

\subsubsection{Geometry dependence, eccentricity and \Np}
\begin{figure*}[htbp]
\includegraphics[width=0.998\linewidth]{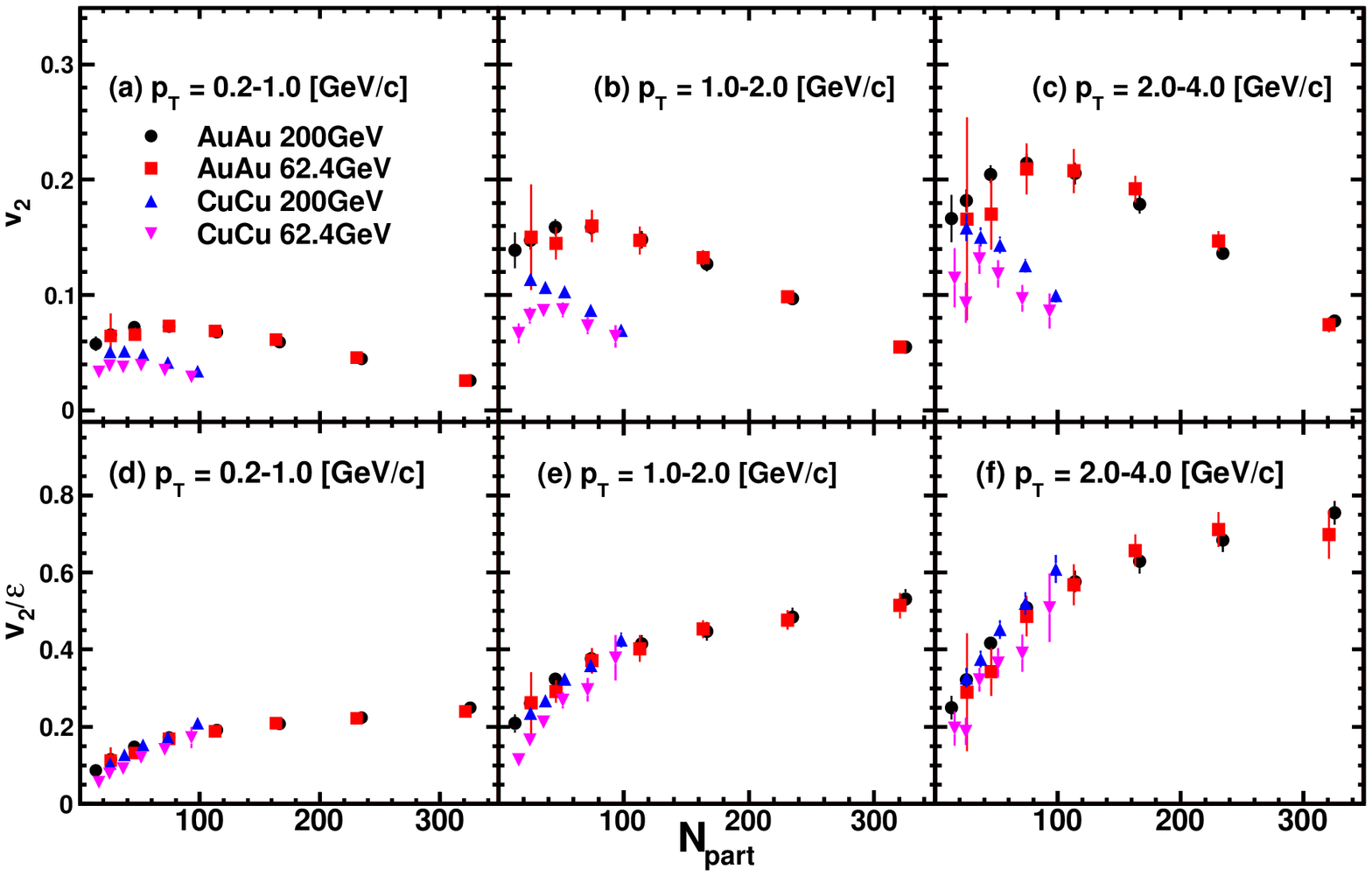}
\caption{\label{fig:v2_npart_4systems} (Color online)
The top three panels show the comparison of integrated \vtwo as a function 
of \Np and the bottom three panels show the comparison of the normalized 
\vtwo /\eps ~vs. \Np in both Au$+$Au and Cu$+$Cu at 200~GeV and 62.4~GeV.  
The ranges of \pt~integration are 0.2--1.0, 1.0--2.0 and 2.0--4.0 
GeV/$c$ from left to right and top to bottom panels respectively. Both 
statistical and systematic uncertainties are included in the error bars. }
\end{figure*}

\begin{figure*}[htbp]
\includegraphics*[width=0.998\linewidth]{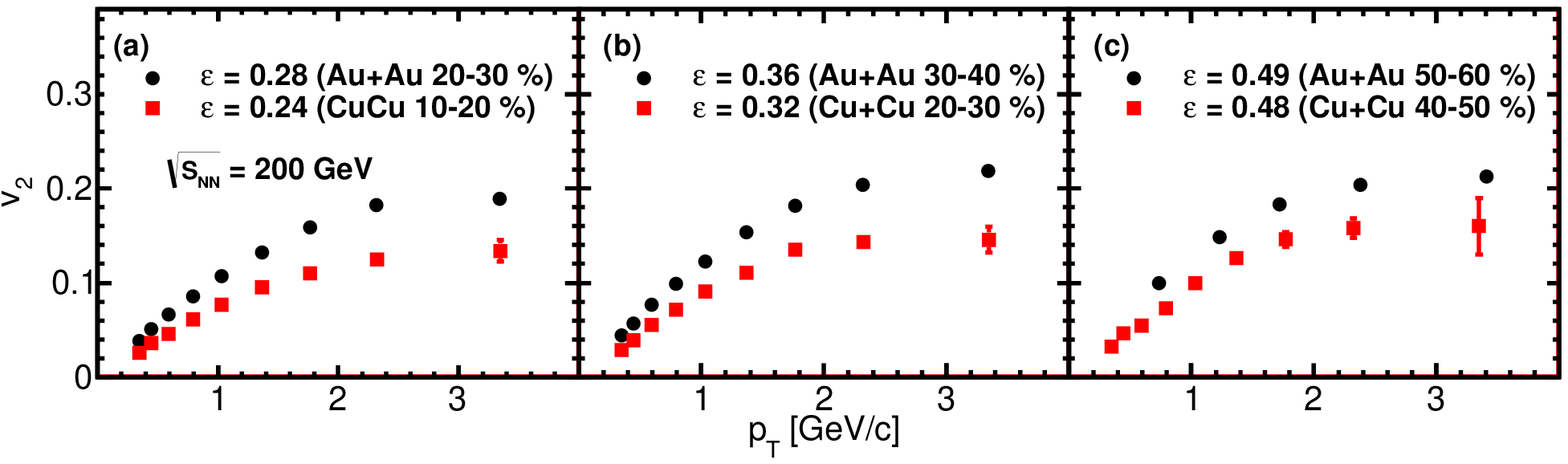}
\caption{\label{fig:auau_cucu200_v2_same_eps} (Color online)
Comparison of \vtwo (\pt ) at 200~GeV for two example systems with 
different collision size (Au$+$Au or Cu$+$Cu) but approximately the same 
\eps.  Black symbols indicate Au$+$Au and red symbols indicate Cu$+$Cu. 
The average number of participants \Np is 166.6 for 20\%--30\%, 114.2 for 
30\%--40\% and 45.5 for 50\%--60\% at Au$+$Au collisions, and \Np is 
73.6 for 10\%--20\%, 53.0 for 20\%--30\% and 25.4 for 40\%--50\% at 
Cu$+$Cu collisions.}
\end{figure*}

There are two ways to establish the extent that \vtwo changes with the 
system size: one is to change the collision centrality, the other is to 
change the colliding nuclei. As seen in 
Fig.~\ref{fig:v2_npart_4systems}, the measured \vtwo in Cu$+$Cu 
collisions is smaller than that of Au$+$Au at a comparable \Np.

Because \eps is different between Au$+$Au and Cu$+$Cu collisions at the 
same \Np, we can try to normalize \vtwo by \eps. In the lower row of 
Fig.~\ref{fig:v2_npart_4systems}, \vtwo normalized by \eps is similar in 
magnitude for both Cu$+$Cu and Au$+$Au collisions. This confirms that the 
eccentricity normalization can account for the effect of the initial 
geometrical anisotropy~\cite{Alver:2006wh}. The exception is that 
the Cu$+$Cu 62.4~GeV data falls below the other data.  Note that the ratio 
$\vtwo /\eps$ also depends on centrality (\Np) and that there is a similar 
rate of increase of $\vtwo /\eps$ with \Np for all three \pt bins: 
0.2--1.0, 1.0--2.0, and 2.0--4.0 GeV/$c$. This pattern suggests the 
need for an additional normalization or scaling factor that depends on 
\Np.

Figure~\ref{fig:auau_cucu200_v2_same_eps} is a comparison of \vtwo as a 
function of \pt for centrality classes that have approximately the same 
value of \eps but with different values of \Np. The average \Np is 166.6 
for 20\%--30\%, 114.2 for 30\%--40\% and 45.5 for 50\%--60\% in Au$+$Au 
collisions, while \Np is 73.6 for 10\%--20\%, 53.0 for 20\%--30\% and 
25.4 for 40\%--50\% in Cu$+$Cu collisions.  It can be clearly seen that 
\vtwo increases with \Np for similar \eps.

\begin{figure*}[htbp]
\includegraphics[width=0.998\linewidth]{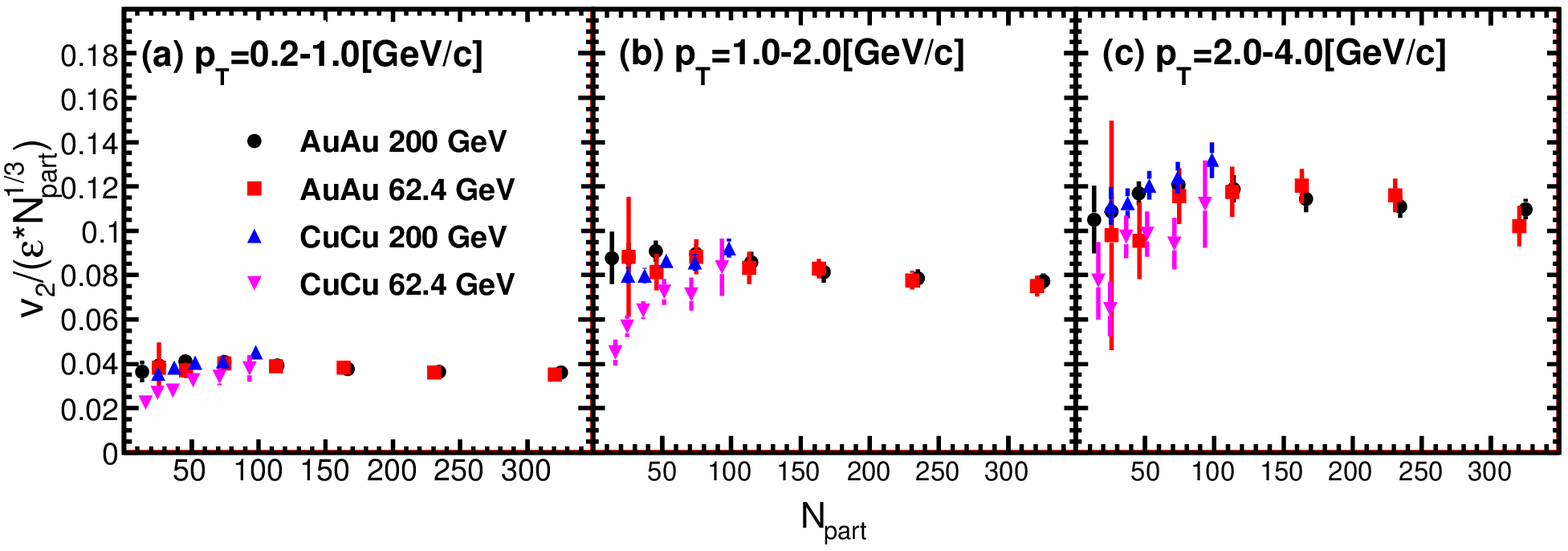}
\caption{ \label{fig:v2_ov_eps_npart1ov3_npart_4systems} (Color online)
Comparison of integrated \vtwo /(\eps $\cdot$ \NpOT ) as a function of \Np 
for two collision energies and two collision systems, Au$+$Au at 200~GeV, 
Au$+$Au at 62.4 GeV, Cu$+$Cu at 200~GeV and Cu$+$Cu at 62.4 GeV.  Ranges 
of \pt integration are 0.2--1.0, 1.0--2.0 and 2.0--4.0~GeV/$c$ from 
left to right panels respectively.  All uncertainties from the measured 
\vtwo, \eps, and \Np are included in the error bars. }
\end{figure*}

\subsubsection{Participant \NpOT scaling}

We empirically explore using \NpOT as a potential scaling factor of \vtwo 
in addition to \eps. We draw on results with a different observable, 
namely that the HBT source sizes at RHIC have been observed to scale with 
\NpOT~\cite{Afanasiev:2009ii}. Under the phenomenological assumption that \Np is 
proportional to the volume of hot/dense matter formed in high-energy 
nuclear collisions, \NpOT can be considered as a quantity proportional to 
a length scale.

Figure~\ref{fig:v2_ov_eps_npart1ov3_npart_4systems} plots \vtwo /(\eps 
$\cdot$ \NpOT ) for integrated bins of \pt = 0.2--1.0, 1.0--2.0, and 
2.0--4.0 GeV/$c$. This combination of two scaling factors works well, 
i.e. the scaled data are at comparable values, with the exception of the 
Cu$+$Cu data at 62.4 GeV which deviate from this scaling, particularly at 
\Np $\leq$ 40. That this empirical $\vtwo/(\eps\cdot\NpOT)$ scaling works 
well suggests that \vtwo is determined by both the initial geometrical 
anisotropy and the number of participants.

Other scalings for the system size dependence have been suggested, 
particularly $1/S_{xy} dN/dy$ \cite{Adler:2002pu} where $S_{xy}$ is the 
transverse area of the participant zone. Because dN/dy is proportional to 
\Np at a given beam energy and $S_{xy}$ is approximately proportional to 
(\Np)$^{2/3}$, $1/S_{xy} dN/dy$ is then proportional to \NpOT.


\section{results for {\vtwo} of identified charged hadrons}

More information can be obtained by examining \vtwo for charged pions, 
kaons and (anti) protons ($\pi$/K/p)  each as a function of transverse 
momentum \pt . The charged particles are identified by TOF and EMCal and 
the data are presented for several classes of collision centrality;

\begin{enumerate}
 \item Au$+$Au collisions at \sqsn = 62.4 GeV  
 \begin{itemize}
     \item 10\%--40\% (Particles and antiparticles are measured separately.) 
     \item 10\% bins from 0\% to 50\%  (Particles and antiparticles are measured together.) 
 \end{itemize}
 \item Au$+$Au collisions at \sqsn = 200 GeV  
 \begin{itemize}
     \item 0\%--92\% (Particles and antiparticles are measured separately.) 
     \item 10\% bins from 0\% to 50\%  (Particles and antiparticles are measured together.) 
 \end{itemize}
 \item Cu$+$Cu collisions at \sqsn = 200 GeV  
 \begin{itemize}
     \item 10\% bins from 0\% to 50\% (Particles and antiparticles are measured together.) 
 \end{itemize}
\end{enumerate}    

Note we do not present Cu$+$Cu 62.4~GeV data in this section because there 
were insufficient statistics to determine \vtwo for identified particles.



\subsection{Beam energy dependence}
\label{subsection:Beam_energy_dependence_PIDv2}

\begin{figure*}[htbp]
\includegraphics[width=0.998\linewidth]{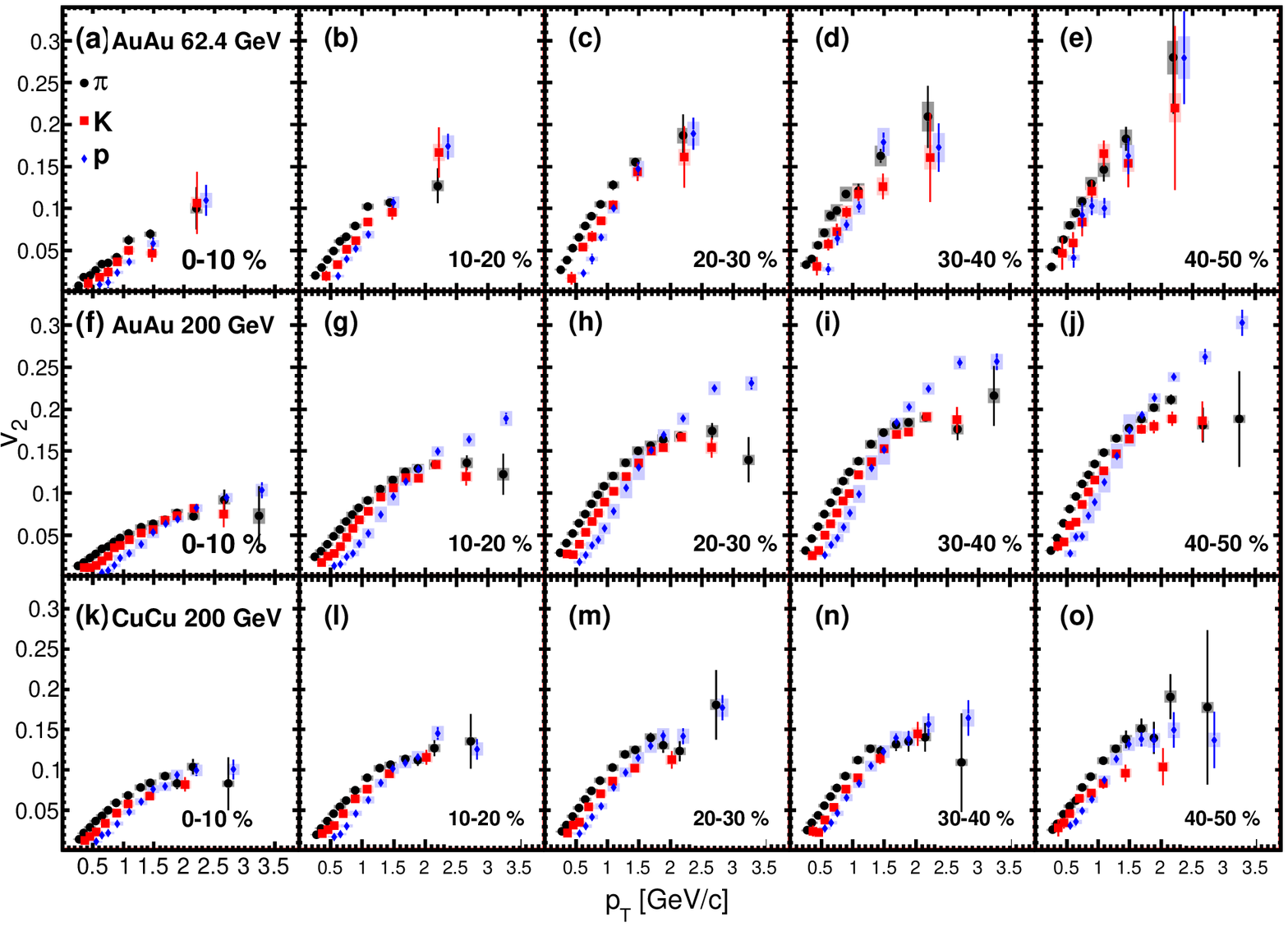}
\caption{\label{fig:pidv2_pt_10step_auau20062cucu200} (Color online)
\vtwo vs. \pt for $\pi/K/p$ emitted from Au$+$Au at 62.4 and 200~GeV and 
Cu$+$Cu at~200~GeV collisions for the centralities indicated. The lines 
for each point indicate the statistical uncertainties, and the boxes are 
systematic uncertainties. In many cases, the systematic uncertainties are smaller than the symbols.}
\end{figure*}


\begin{figure*}[htbp]	
\includegraphics[width=0.998\linewidth]{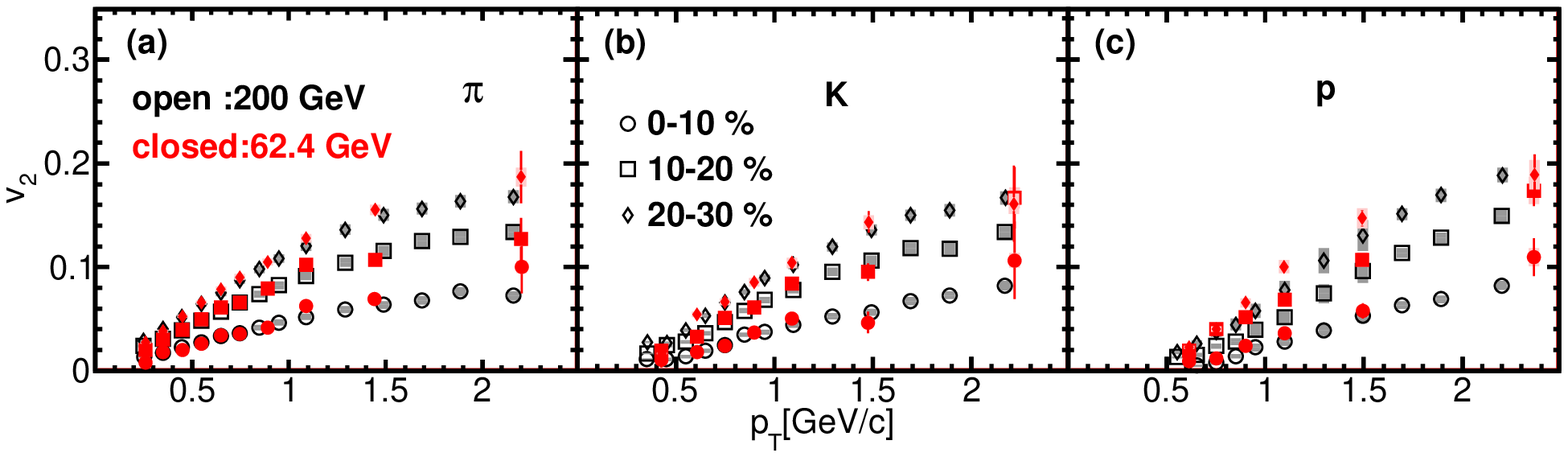}
\caption{\label{fig:pidv2pt_20to30_auau200_62} (Color online)
Comparison of \vtwo between \sqsn = 62.4 and 200~GeV for $\pi$/K/p emitted 
from 0\%--10\%, 10\%--20\% and 20\%--30\% central Au$+$Au collisions. Both 
results for all species agree within the errors.  The lines indicate the 
statistical uncertainties at each point and the boxes indicate the 
systematic uncertainties. In many cases, the systematic uncertainties are smaller than the symbols.}
\end{figure*}

Figure~\ref{fig:pidv2_pt_10step_auau20062cucu200} shows a summary of \vtwo 
measurements of identified particles $\pi/K/p$ for three different data 
sets; Au$+$Au at 62.4 and 200~GeV and Cu$+$Cu at 200~GeV. 
Figure~\ref{fig:pidv2pt_20to30_auau200_62} shows the comparison between 
62.4 and 200~GeV for Au$+$Au collisions. The measured \vtwo in the 62.4 
and 200~GeV data sets are consistent, within the systematic uncertainties, 
with the exception of proton \vtwo at 62.4 GeV which is slightly higher 
than at 200~GeV in the lower \pt region.  These small differences could be 
caused by larger radial flow at higher \sqsn, especially for heavier 
particles such as protons.

The observation that the proton \vtwo is larger at 62.4 GeV than at 200 GeV 
for Au$+$Au collisions is opposite to the earlier observation that 
inclusive charged \vtwo at 62.4~GeV is lower than that at 200~GeV Cu$+$Cu.  
Therefore, the differences in lower \vtwo for inclusive charged hadrons 
from Cu$+$Cu may be caused by different physics than the radial flow 
effect seen in Au$+$Au collisions.

\subsection{Particle-antiparticle comparison}

\begin{figure*}[htbp]
\includegraphics[width=0.998\linewidth]{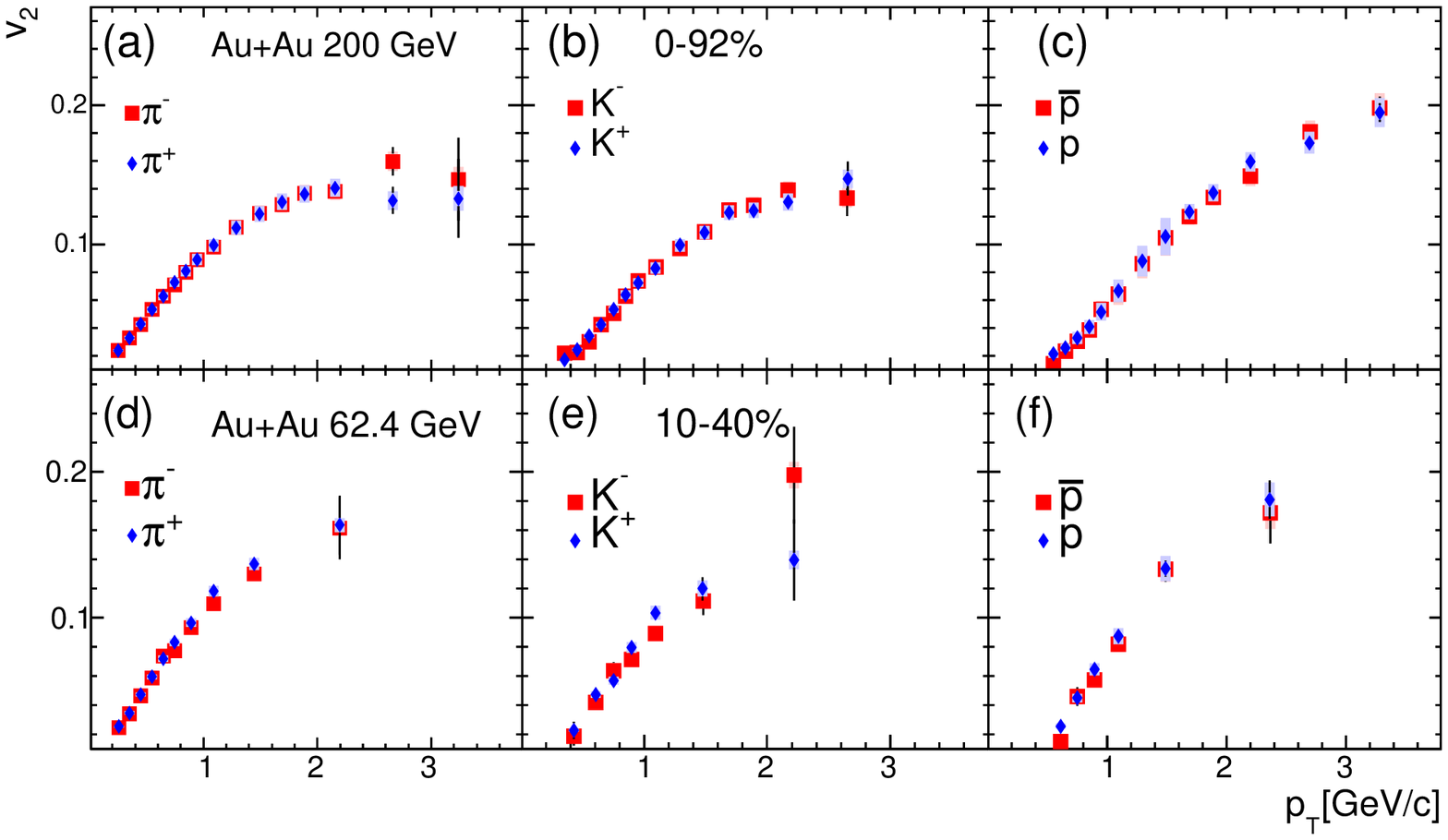}
\caption{\label{fig:pidv2pt_MB_auau200_10to40_auau62} (Color online)
Comparison of the \vtwo of particles, antiparticles, for a minimum bias 
sample 0\%--92\% at 200~GeV and 10\%--40\% central at 62.4 GeV in Au$+$Au 
collisions. The lines for each point indicate the statistical 
uncertainties, and the boxes are systematic uncertainties. In many cases, the systematic uncertainties are smaller than the symbols.} 
\includegraphics[width=0.998\linewidth]{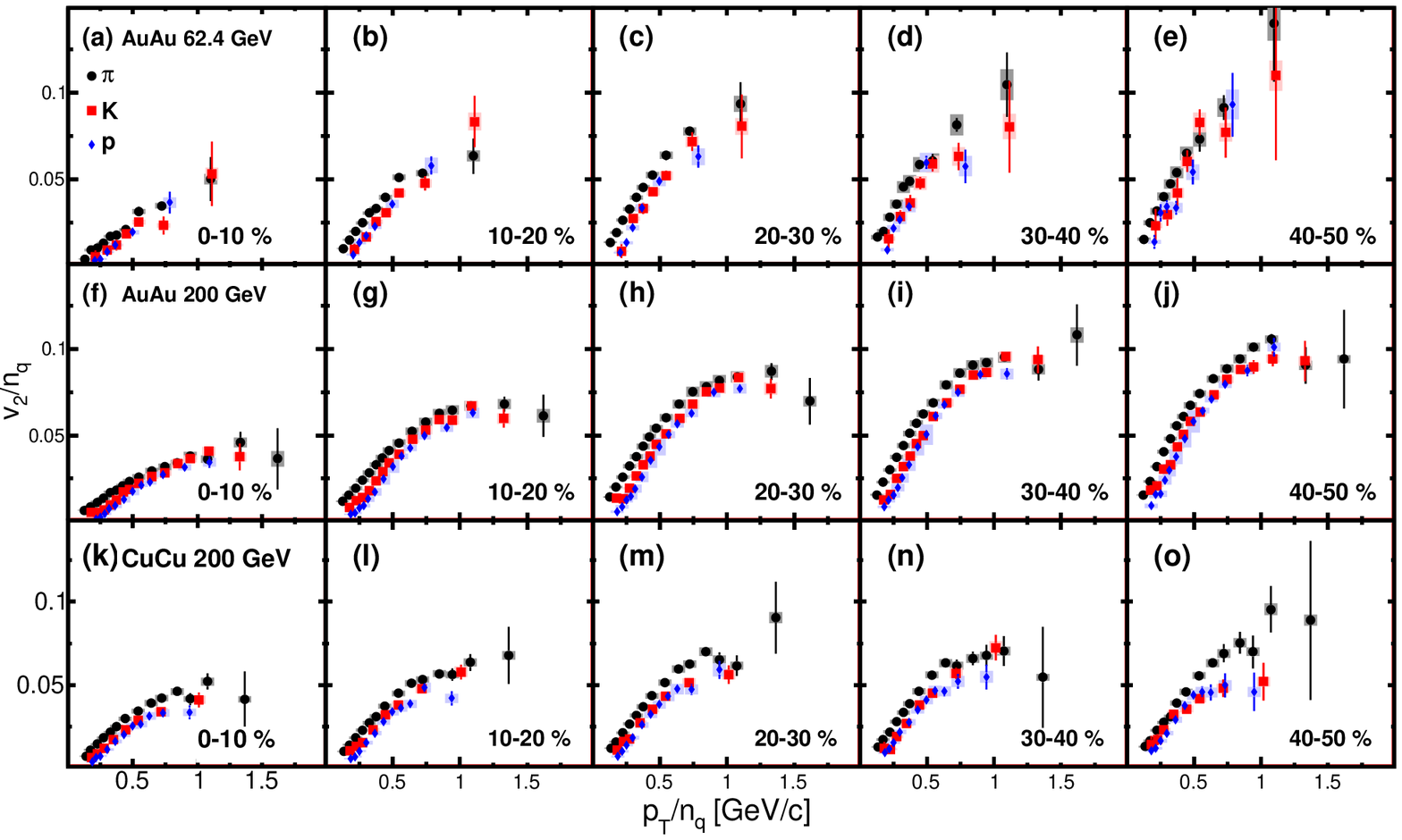}
\caption{\label{fig:pidv2nq_ptnq_10step_auau20062cucu200} (Color online)
The ratio \vtwo/\nq vs. \pt/\nq for $\pi/K/p$ emitted from Au$+$Au at 62.4 
and 200~GeV and Cu$+$Cu at 200~GeV collisions for the centralities 
indicated. The lines for each point indicate the statistical 
uncertainties, and the boxes are systematic uncertainties. In many cases, the systematic uncertainties are smaller than the symbols.}
\end{figure*}

When we examine identified \vtwo we will combine opposite charged 
particles, e.g. $\pi^{\pm}$, to form $\pi$ \vtwo. Prior results on the 
ratio of \vtwo for antiparticles and particles can be found in 
Refs.~\cite{Abelev:2007qg, Adamczyk:2013gw}. In this section we compare 
the particle and antiparticle \vtwo in Au$+$Au collisions at 200 and 
62.4~GeV in wide centrality classes: a minimum bias sample (0\%--92\%) for 
200~GeV and 10\%--40\% for 62.4~GeV data.  The first and second rows of 
plots in Fig.~\ref{fig:pidv2pt_MB_auau200_10to40_auau62} present \vtwo as 
a function of \pt for $\pi^{\pm}$, K$^{\pm}$, p and $\bar{\textrm{p}}$ in 
Au$+$Au collisions at 200 and 62.4~GeV. The lines for each point are the 
statistical uncertainties and the boxes are systematic uncertainties.
 
At both 200 and 62.4~GeV, the the measured Au$+$Au \vtwo values of 
particle and antiparticle are comparable to each other within uncertainty, 
though there is a possible indication of a small reduction of anti-proton 
\vtwo at lower \pt. When we combine particle and anti-particle \vtwo we 
average over these differences.

\begin{figure*}[htbp]	
\includegraphics[width=0.998\linewidth]{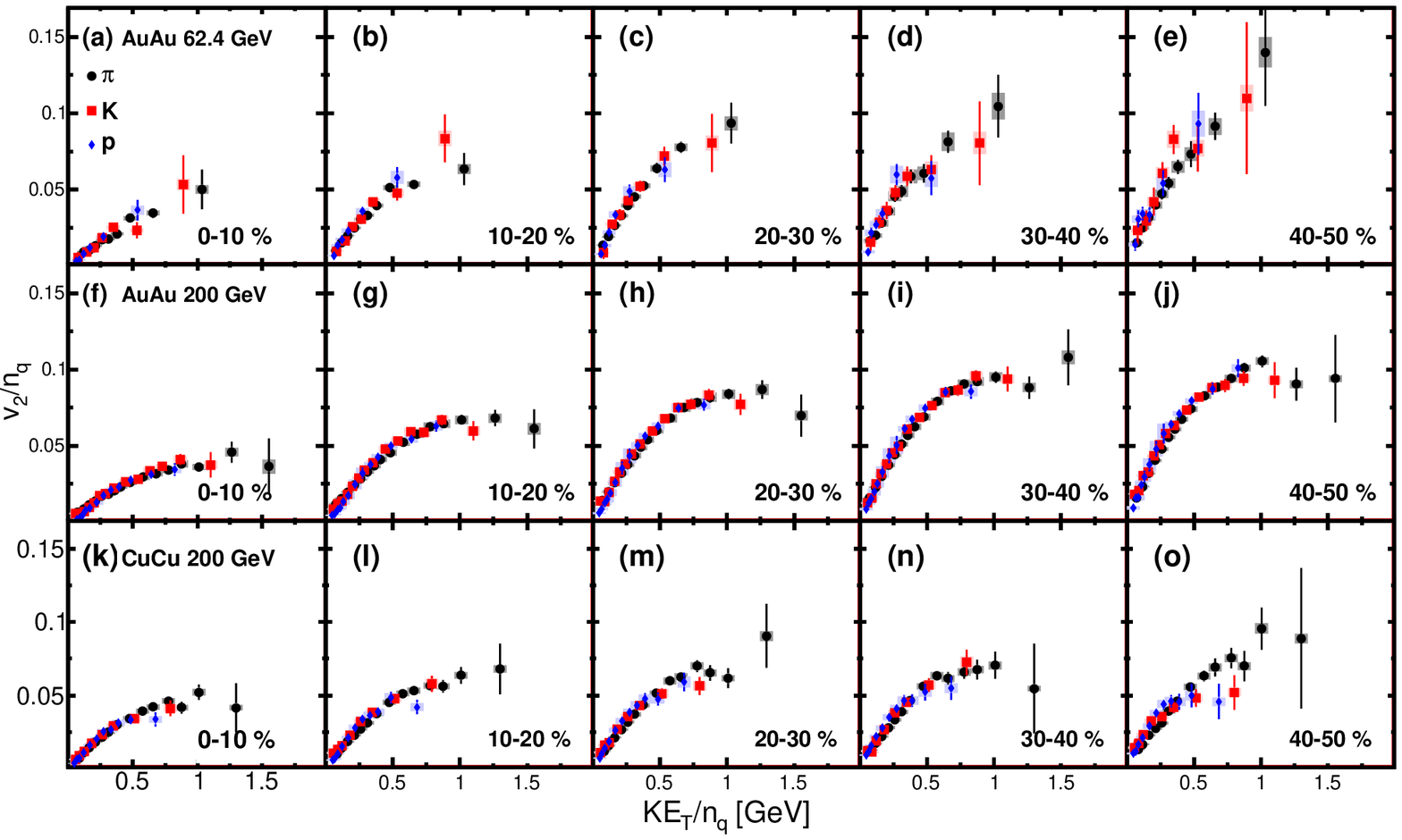}
\caption{\label{fig:pidv2nq_ketnq_10step_auau20062cucu200} (Color online)
The ratio \vtwo/\nq vs. \KET/\nq for $\pi/K/p$ emitted from Au$+$Au at 
62.4 and 200~GeV and Cu$+$Cu at 200~GeV collisions for the centralities 
indicated. The lines for each point indicate the statistical uncertainties 
and the boxes are systematic uncertainties. In many cases, the systematic uncertainties are smaller than the symbols.}
\includegraphics[width=0.998\linewidth]{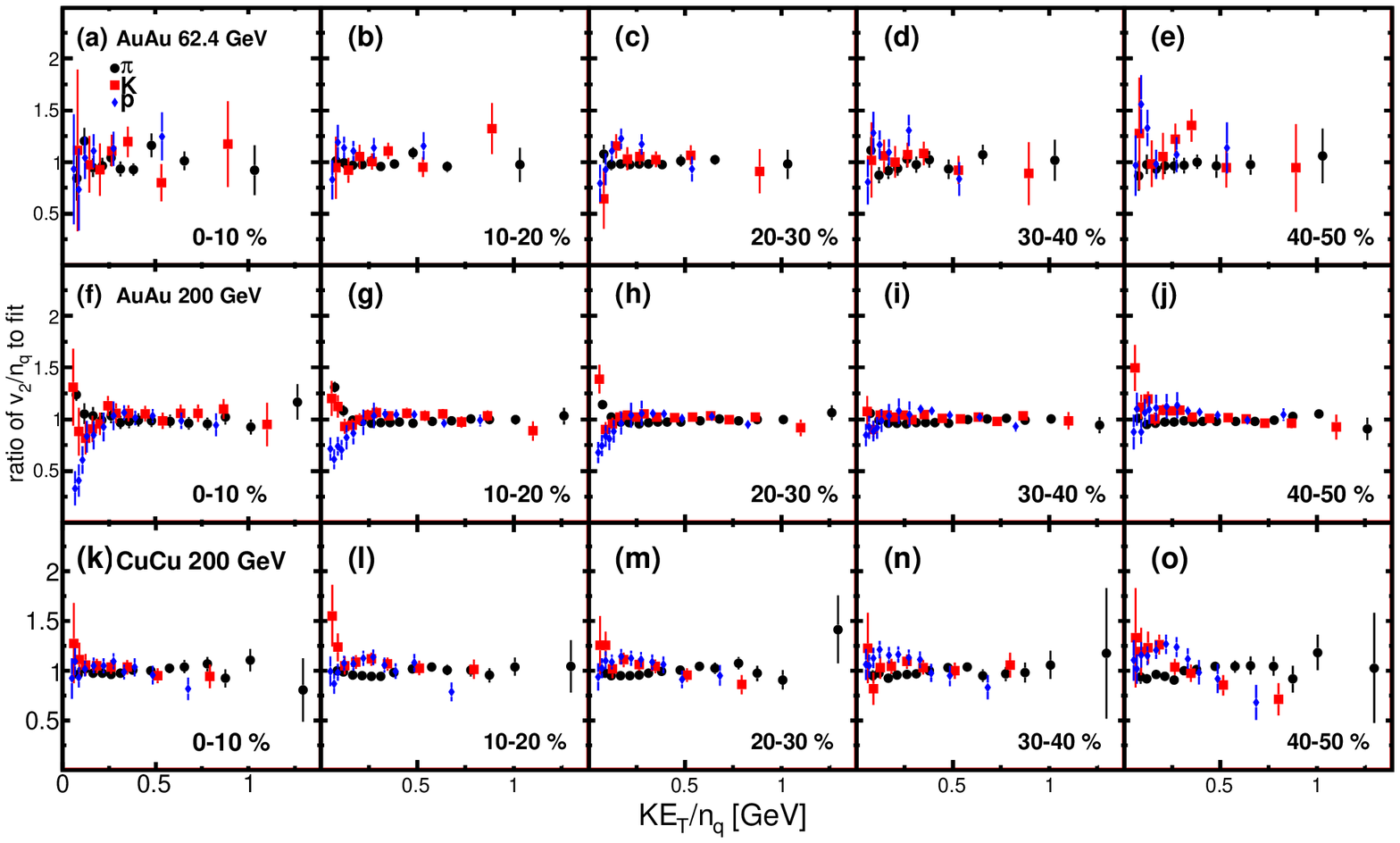}
\caption{\label{fig:pidratio_v2nq_ketnq_10step_auau20062cucu200} (Color online)
The ratio of \vtwo/\nq vs. \KET/\nq to the fit for $\pi/K/p$ emitted from 
Au$+$Au at 62.4 and 200~GeV and Cu$+$Cu at 200~GeV collisions for the 
centralities indicated. The lines for each point indicate the statistical 
uncertainties.}
\end{figure*}

\subsection{Number of valence quark \nq scaling of \vtwo}
\label{section:nq_scaling}

The \vtwo measurements of identified particles $\pi/K/p$ for three 
different data sets; Au$+$Au at 62.4 and 200~GeV and Cu$+$Cu at 200~GeV 
collisions are re-plotted in 
Fig.~\ref{fig:pidv2nq_ptnq_10step_auau20062cucu200} after scaling by the 
number of constituent quarks for both \vtwo and \pt axes as shown. An 
alternative scaling is to use transverse kinetic energy. We define 
transverse kinetic energy as \KET= \mT$-m$, where m is the mass of the 
hadron and \mT$=\sqrt{p_{T}^2 + m^{2}}$. The quark number scaled \vtwo are 
shown as a function of \KET/\nq for all three data sets in 
Fig.~\ref{fig:pidv2nq_ketnq_10step_auau20062cucu200}.

Note that at higher values, \KET/\nq $>$ 0.7, PHENIX has observed 
significant deviations from \nq scaling for Au$+$Au noncentral 
collisions\cite{Adare:2012vq}. Those higher \KET results indicate that the 
azimuthal anisotropy of these high \KET particles are impacted by 
mechanisms such as parton-energy loss, jet chemistry, and/or different 
fragmentation functions. For comparison, at the 
LHC~\cite{Abelev:2012di,Abelev:2014pua}, 
\vtwo does not scale well with the quark number and 
transverse kinetic energy of the hadron in any range of \KET/\nq, with up 
to 40\% deviations observed at low values of \KET/\nq.

To quantify how well the number of quark scaling with \KET works with the 
current data, we fit all the hadron species data in 
Figure~\ref{fig:pidv2nq_ketnq_10step_auau20062cucu200} with a common 
polynomial function for each centrality and colliding system. We divide 
the data by these fits to compare how close different hadron species are 
to the common scaled shape of \vtwo. 
Figure~\ref{fig:pidratio_v2nq_ketnq_10step_auau20062cucu200} shows these 
ratios as a function of \KET/\nq for $\pi$/K/p in Au$+$Au and Cu$+$Cu. 
Deviations from the fitted polynomial function are observed, especially 
with the high statistics data sets at 200~GeV Au$+$Au and 200~GeV Cu$+$Cu 
collisions. For Au$+$Au central collisions in the low \KET/\nq region 
(\KET/\nq $< 0.1$~GeV), protons sit below the common scaling fit and rise 
above the fit at moderate \KET/\nq. These deviations systematically change 
with centrality, i.e. the proton \vtwo is smaller than pion \vtwo at low 
\KET/\nq in the most central Au$+$Au collisions at 200~GeV, while the 
proton \vtwo becomes larger than pion \vtwo in peripheral collisions. The 
proton \vtwo is also larger than the pion \vtwo at low \KET/\nq in 200~GeV 
Cu$+$Cu peripheral collisions. The proton and pion \vtwo become comparable 
in central Cu$+$Cu collisions.  It is noted that the location where the 
proton and pion \vtwo flows are comparable occurs at a similar number of 
participants \Np for Au$+$Au and Cu$+$Cu. This could be explained by an 
increase in radial flow as a function of the number of participants, which 
effectively reduces the proton \vtwo relative to the pion \vtwo for a 
given \pt \cite{Voloshin:2008dg}.

For Cu$+$Cu collisions at 200~GeV, the bottom five panels of 
Figs.~\ref{fig:pidv2nq_ptnq_10step_auau20062cucu200} 
and~\ref{fig:pidv2nq_ketnq_10step_auau20062cucu200} show the \vtwo/\nq vs. 
\pt/\nq and \KET/\nq, respectively for $\pi$/K/p emitted from Cu$+$Cu 
collisions at 200~GeV for the five centrality bins: 0\%--10\%, 10\%--20\%, 
20\%--30\%, 30\%--40\% and 40\%--50\%. For the smaller system of Cu$+$Cu 
at 200~GeV (the bottom row of 
Fig.~\ref{fig:pidv2nq_ketnq_10step_auau20062cucu200}), quark number with 
\KET scalings reduces the spread in \vtwo values better than \pt scaling 
in Fig.~\ref{fig:pidv2nq_ptnq_10step_auau20062cucu200}, especially for the 
more central collisions between 0\%--40\%.  For peripheral Cu$+$Cu 
collisions, the number of quark scaling with \KET does not work well. The 
deviation from \nq scaling seems to be largest at peripheral collisions, 
i.e. at 40\%--50\%, especially between pions and protons.

\begin{figure*}[htbp]	
\includegraphics[width=0.998\linewidth]{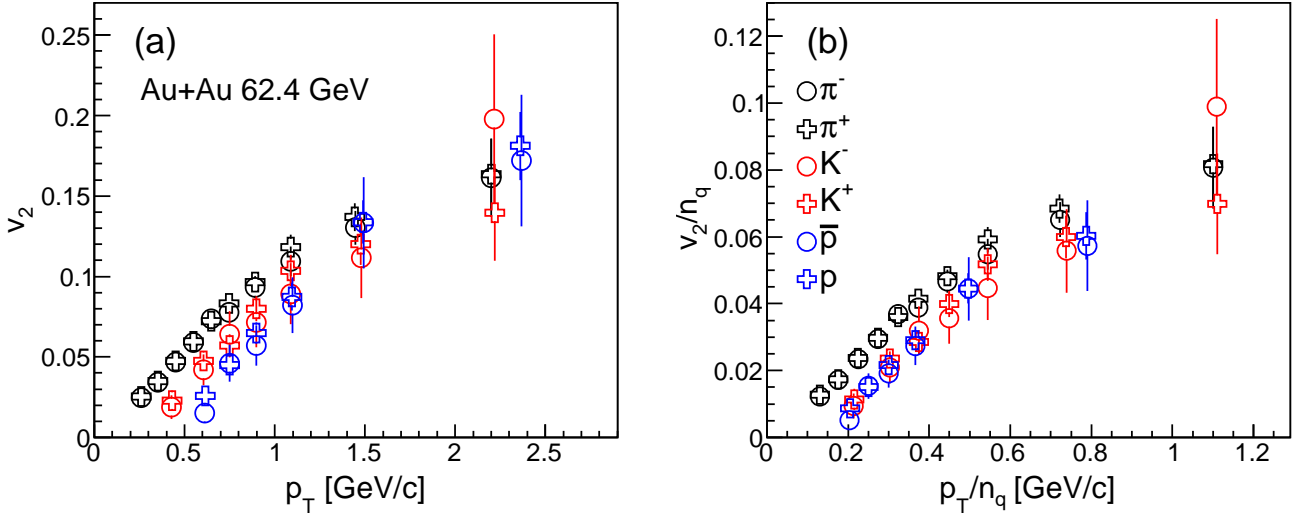}
\caption{\label{fig:pidv2pt_10to40_scaling_auau62} (Color online)
The left panel shows \vtwo vs. \pt, the right panel is the ratio \vtwo 
/\nq vs. \pt /\nq for the indicated hadrons emitted from 10\%--40 \% 
central Au$+$Au collisions in Au$+$Au at 62.4~GeV. The error bars include 
both systematic and statistical uncertainties.}
\end{figure*}

\begin{figure}[htbp]	
\includegraphics[width=1.0\linewidth]{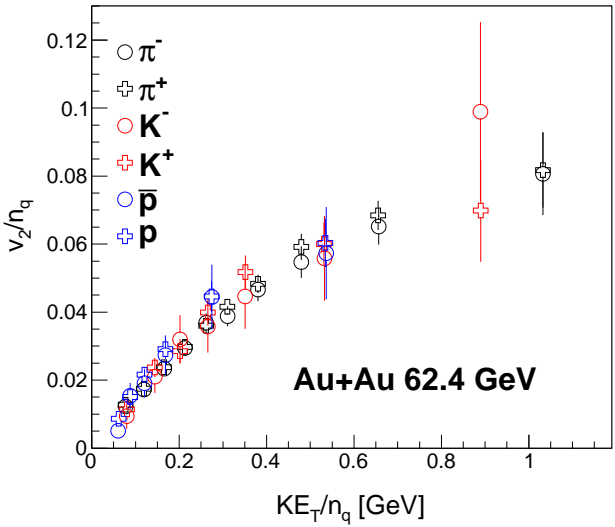}
\caption{\label{fig:pidv2ket_10to40_scaling_auau62} (Color online)
The ratio \vtwo /\nq vs. \KET /\nq for the indicated hadrons emitted from 
10\%-40\% central Au$+$Au collisions at 62.4~GeV. The error 
bars include both systematic and statistical uncertainties.}
\end{figure}

We examine in more detail the scaling at low \KET in the 62.4~GeV data in 
stages. First, the left panel in 
Fig.~\ref{fig:pidv2pt_10to40_scaling_auau62} summarizes the unscaled \vtwo 
data from 10\%--40\% central Au$+$Au collisions at 62.4~GeV. The \vtwo 
values are broadly spread in their magnitude.  A reduction in spread is 
observed in the right panel when \nq, the number of valence quarks, is 
used as a scaling. However the scaled \vtwo values do not collapse to a 
universal curve. Figure~\ref{fig:pidv2ket_10to40_scaling_auau62} does show 
a better scaling with \KET/\nq.

Overall, the combined \nq$-$~\KET scaling works well (typical deviations 
less than 20\%) for $0.1<$\KET/\nq$<$1~GeV, indicating that the elliptic 
collective motion is created at a level consistent with constituent quarks 
both at 62.4~GeV in Au$+$Au and at 200~GeV in Cu$+$Cu.

\subsection{Universal \vtwo scaling }

\begin{figure*}[htbp]
\includegraphics[width=0.998\linewidth]{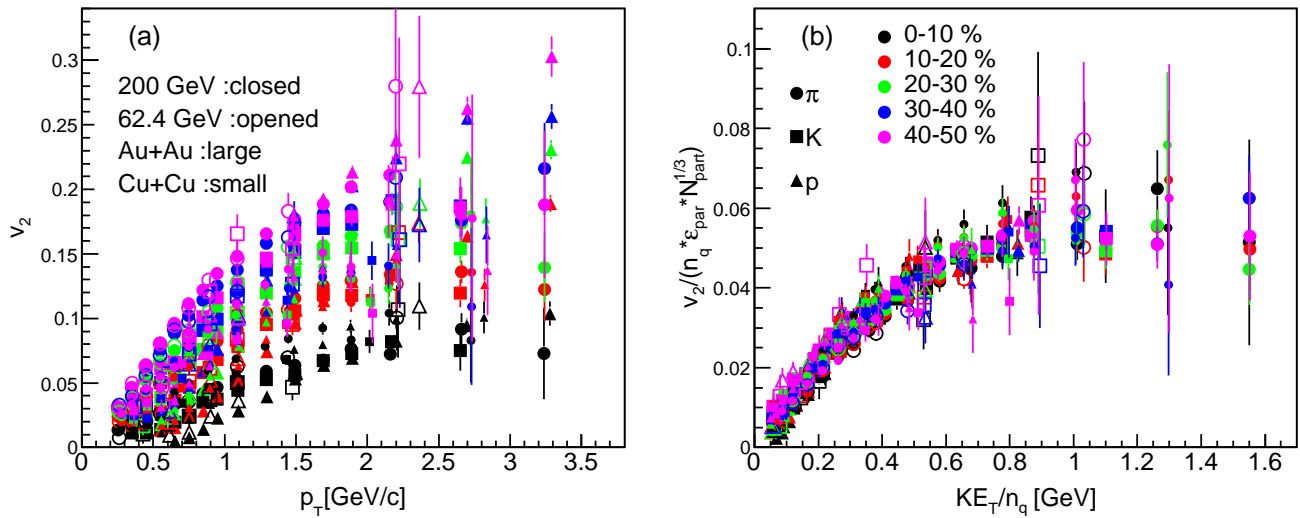}
\caption{\label{v2_Npart1ov3_universal_45} (Color online)
The left panel shows \vtwo vs. \pt and the right panel shows  \vtwo 
/(\eps $\cdot$ \NpOT $\cdot$ \nq ) vs. \KET /\nq for $\pi$/K/p in 
Au$+$Au at 200~GeV, in Au$+$Au at 62.4 GeV and in Cu$+$Cu 
at 200~GeV for five centrality bins over 0\%--50\% in 10\% steps for  
each system. There are 45 data sets in each panel.}
\end{figure*}

We consider a universal \vtwo scaling for all the \vtwo measurements in 
this paper for identified hadrons between $0.1 <~$\KET/\nq$<$ 1~GeV.  
Within a given collision system, i.e. each centrality bin for each set of 
Au$+$Au and Cu$+$Cu collisions, we first apply quark number \nq scaling 
and \KET scaling. Then we apply the eccentricity normalization and \NpOT 
scaling for each colliding system.  Because we have observed that \vtwo 
saturates with beam energy between 62 -200~GeV, we do not apply any 
scaling with beam energy.  The \vtwo data with the four factors applied 
(quark number scaling, \KET scaling, eccentricity normalization and \NpOT 
scaling) are shown as a function of \KET /\nq in 
Fig.~\ref{v2_Npart1ov3_universal_45}, which includes data from 
Au$+$Au at 200~GeV, Au$+$Au at 62.4 GeV and Cu$+$Cu at 200~GeV at five 
centrality bins over 0\%--50\% in 10\% steps for each system.  There are 
45 \vtwo data sets in total. The combined data is fit with a single 
3rd-order-polynomial, producing a $\chi ^{2}/NDF$ = 1034/490 = 2.11 
(including both statistical and systematic uncertainties).  Note there is 
no Cu$+$Cu 62.4~GeV data in Fig.~\ref{v2_Npart1ov3_universal_45}, because 
there were insufficient statistics to determine \vtwo for identified particles.

If we apply the \NcollOT scaling to the same data sets instead of \NpOT 
scaling, we obtain $\chi ^{2}/NDF=2643/490=5.39$. Therefore, \NpOT is a 
better scaling factor than \NcollOT. As we mentioned 
Section~\ref{section:nq_scaling}, there are some deviations from the quark 
number and \KET scalings, therefore this \NpOT normalized curve is not 
perfectly a single line.  Further investigation of these deviations would 
require higher precision measurements.

\section{summary and conclusion}

We have measured the strength of the elliptic anisotropy, \vtwo , for 
inclusive charged hadrons and identified charged hadrons ($\pi$/K/p) in 
Au$+$Au and Cu$+$Cu collisions at \sqsn = 200 and 62.4 GeV to study the 
dependence of \vtwo on collision energy, species and centrality.  Results 
of this systematic study reveal the following features.  Comparisons 
between 200 and 62.4~GeV collisions demonstrate that \vtwo as a function 
of \pt does not depend on beam energy in Au$+$Au. In Cu$+$Cu, the \vtwo at 
62.4~GeV is slightly lower than that at 200~GeV.

One possibility for the lower \vtwo values 62.4 GeV in Cu$+$Cu is less 
complete thermalization in small systems at lower beam energies. At least 
two types of theoretical models have been used to investigate the question 
of incomplete thermalization for systems formed at RHIC. Borghini argues 
that because \vtwo/\eps depends on dN/dy ~\cite{Borghini:2005hx}, the systems 
formed at RHIC are not fully thermalized during the time when \vtwo 
develops. Borghini argues that this dN/dy dependence can be interpreted as 
dependence on a Knudsen number representing incomplete thermalization. 
Recent hydrodynamical models that include shear viscosity and initial 
fluctuations~\cite{Niemi:2012ry,Song:2011hk,Soltz:2012rk} 
effectively include nonequilibrium effects through the finite viscosity. 
Using a different non-equilibrium approach, microscopic transport models~\cite{Uphoff:2014cba} solve the relativistic Boltzmann equation. 
Both the viscous hydrodynamical and 
the Boltzmann transport models can be tested with our two observation that 
the \vtwo at Cu$+$Cu at 62.4~GeV is slightly lower than that at 200~GeV, 
and that the measured universal scaling breakdowns in peripheral Cu$+$Cu.
    
For various hadron species the measured \vtwo results as a function of \pt 
are well scaled by quark number.   Interestingly, it appears that this
scaling holds also for higher orders in azimuthal anisotropy~\cite{Adare:2014kci}.
The \KET scaling performs better than \pt 
scaling, particularly in the intermediate transverse momentum region (\pt 
= 1--4~GeV/$c$). This scaling property suggests that the matter flows with 
quark-like degrees of freedom, and therefore is consistent with the 
formation of QGP matter~\cite{Adare:2006ti}.  A small deviation 
from \KET scaling can be seen for both Au$+$Au and Cu$+$Cu collisions, and 
this deviation depends on the number of participants \Np. This deviation 
might indicate a restricted region where \KET scaling works well, possibly 
dependent on the strength of the radial flow.

For both Au$+$Au to Cu$+$Cu collisions, we confirm that \vtwo can be 
normalized by participant eccentricity 
(\eps)~\cite{Alver:2006wh}. This indicates that the effect of 
initial geometrical anisotropy can be partially removed by eccentricity 
normalization.  However, \vtwo normalized by \eps still depends on \Np, 
\vtwo is not fully determined by \eps alone and we have empirically found 
that \vtwo /\eps is proportional to \NpOT. The initial participant size 
\NpOT, is related to a length scale or an expansion time scale. Taking 
account all scalings and normalization, the data ``\vtwo /\nq /\eps /\NpOT 
vs. \KET /\nq'' lie on a universal curve for $0.1<~$\KET/\nq$<1$~GeV.

 
\section*{ACKNOWLEDGMENTS}   

We thank the staff of the Collider-Accelerator and Physics
Departments at Brookhaven National Laboratory and the staff of
the other PHENIX participating institutions for their vital
contributions.  We acknowledge support from the 
Office of Nuclear Physics in the 
Office of Science of the Department of Energy, 
the National Science Foundation, 
Abilene Christian University Research Council, 
Research Foundation of SUNY, 
and Dean of the College of Arts and Sciences, Vanderbilt University 
(U.S.A),
Ministry of Education, Culture, Sports, Science, and Technology
and the Japan Society for the Promotion of Science (Japan),
Conselho Nacional de Desenvolvimento Cient\'{\i}fico e
Tecnol{\'o}gico and Funda\c c{\~a}o de Amparo {\`a} Pesquisa do
Estado de S{\~a}o Paulo (Brazil),
Natural Science Foundation of China (P.~R.~China),
Ministry of Education, Youth and Sports (Czech Republic),
Centre National de la Recherche Scientifique, Commissariat
{\`a} l'{\'E}nergie Atomique, and Institut National de Physique
Nucl{\'e}aire et de Physique des Particules (France),
Bundesministerium f\"ur Bildung und Forschung, Deutscher
Akademischer Austausch Dienst, and Alexander von Humboldt Stiftung (Germany),
National Science Fund, OTKA, K\'aroly R\'obert University College,
and the Ch. Simonyi Fund (Hungary),
Department of Atomic Energy (India), 
Israel Science Foundation (Israel), 
National Research Foundation and WCU program of the 
Ministry Education Science and Technology (Korea),
Ministry of Education and Science, Russian Academy of Sciences,
Federal Agency of Atomic Energy (Russia),
VR and Wallenberg Foundation (Sweden), 
the U.S. Civilian Research and Development Foundation for the
Independent States of the Former Soviet Union, 
the US-Hungarian NSF-OTKA-MTA, 
and the US-Israel Binational Science Foundation.


\begin{thebibliography}{43}%
\makeatletter
\providecommand \@ifxundefined [1]{%
 \@ifx{#1\undefined}
}%
\providecommand \@ifnum [1]{%
 \ifnum #1\expandafter \@firstoftwo
 \else \expandafter \@secondoftwo
 \fi
}%
\providecommand \@ifx [1]{%
 \ifx #1\expandafter \@firstoftwo
 \else \expandafter \@secondoftwo
 \fi
}%
\providecommand \natexlab [1]{#1}%
\providecommand \enquote  [1]{``#1''}%
\providecommand \bibnamefont  [1]{#1}%
\providecommand \bibfnamefont [1]{#1}%
\providecommand \citenamefont [1]{#1}%
\providecommand \href@noop [0]{\@secondoftwo}%
\providecommand \href [0]{\begingroup \@sanitize@url \@href}%
\providecommand \@href[1]{\@@startlink{#1}\@@href}%
\providecommand \@@href[1]{\endgroup#1\@@endlink}%
\providecommand \@sanitize@url [0]{\catcode `\\12\catcode `\$12\catcode
  `\&12\catcode `\#12\catcode `\^12\catcode `\_12\catcode `\%12\relax}%
\providecommand \@@startlink[1]{}%
\providecommand \@@endlink[0]{}%
\providecommand \url  [0]{\begingroup\@sanitize@url \@url }%
\providecommand \@url [1]{\endgroup\@href {#1}{\urlprefix }}%
\providecommand \urlprefix  [0]{URL }%
\providecommand \Eprint [0]{\href }%
\providecommand \doibase [0]{http://dx.doi.org/}%
\providecommand \selectlanguage [0]{\@gobble}%
\providecommand \bibinfo  [0]{\@secondoftwo}%
\providecommand \bibfield  [0]{\@secondoftwo}%
\providecommand \translation [1]{[#1]}%
\providecommand \BibitemOpen [0]{}%
\providecommand \bibitemStop [0]{}%
\providecommand \bibitemNoStop [0]{.\EOS\space}%
\providecommand \EOS [0]{\spacefactor3000\relax}%
\providecommand \BibitemShut  [1]{\csname bibitem#1\endcsname}%
\let\auto@bib@innerbib\@empty
\bibitem [{\citenamefont {Adcox}\ \emph {et~al.}(2005)\citenamefont {Adcox}
  \emph {et~al.}}]{Adcox:2004mh}%
  \BibitemOpen
  \bibfield  {author} {\bibinfo {author} {\bibfnamefont {K.}~\bibnamefont
  {Adcox}} \emph {et~al.} (\bibinfo {collaboration} {PHENIX Collaboration}),\
  }\bibfield  {title} {\enquote {\bibinfo {title} {{Formation of dense partonic
  matter in relativistic nucleus-nucleus collisions at RHIC: Experimental
  evaluation by the PHENIX Collaboration}},}\ }\href {\doibase
  10.1016/j.nuclphysa.2005.03.086} {\bibfield  {journal} {\bibinfo  {journal}
  {Nucl. Phys. A}\ }\textbf {\bibinfo {volume} {757}},\ \bibinfo {pages} {184}
  (\bibinfo {year} {2005})}\BibitemShut {NoStop}%
\bibitem [{\citenamefont {Adams}\ \emph {et~al.}(2005)\citenamefont {Adams}
  \emph {et~al.}}]{Adams:2005dq}%
  \BibitemOpen
  \bibfield  {author} {\bibinfo {author} {\bibfnamefont {J.}~\bibnamefont
  {Adams}} \emph {et~al.} (\bibinfo {collaboration} {STAR Collaboration}),\
  }\bibfield  {title} {\enquote {\bibinfo {title} {{Experimental and
  theoretical challenges in the search for the quark gluon plasma: The STAR
  Collaboration's critical assessment of the evidence from RHIC collisions}},}\
  }\href {\doibase 10.1016/j.nuclphysa.2005.03.085} {\bibfield  {journal}
  {\bibinfo  {journal} {Nucl. Phys. A}\ }\textbf {\bibinfo {volume} {757}},\
  \bibinfo {pages} {102} (\bibinfo {year} {2005})}\BibitemShut {NoStop}%
\bibitem [{\citenamefont {Back}\ \emph {et~al.}(2005)\citenamefont {Back} \emph
  {et~al.}}]{Back:2004je}%
  \BibitemOpen
  \bibfield  {author} {\bibinfo {author} {\bibfnamefont {B.~B.}\ \bibnamefont
  {Back}} \emph {et~al.} (\bibinfo {collaboration} {PHOBOS Collaboration}),\
  }\bibfield  {title} {\enquote {\bibinfo {title} {{The PHOBOS perspective on
  discoveries at RHIC}},}\ }\href {\doibase 10.1016/j.nuclphysa.2005.03.084}
  {\bibfield  {journal} {\bibinfo  {journal} {Nucl. Phys. A}\ }\textbf
  {\bibinfo {volume} {757}},\ \bibinfo {pages} {28} (\bibinfo {year}
  {2005})}\BibitemShut {NoStop}%
\bibitem [{\citenamefont {Arsene}\ \emph {et~al.}(2005)\citenamefont {Arsene}
  \emph {et~al.}}]{Arsene:2004fa}%
  \BibitemOpen
  \bibfield  {author} {\bibinfo {author} {\bibfnamefont {I.}~\bibnamefont
  {Arsene}} \emph {et~al.} (\bibinfo {collaboration} {BRAHMS Collaboration}),\
  }\bibfield  {title} {\enquote {\bibinfo {title} {{Quark gluon plasma and
  color glass condensate at RHIC? The Perspective from the BRAHMS
  experiment}},}\ }\href {\doibase 10.1016/j.nuclphysa.2005.02.130} {\bibfield
  {journal} {\bibinfo  {journal} {Nucl. Phys. A}\ }\textbf {\bibinfo {volume}
  {757}},\ \bibinfo {pages} {1} (\bibinfo {year} {2005})}\BibitemShut {NoStop}%
\bibitem [{\citenamefont {Adler}\ \emph {et~al.}(2003)\citenamefont {Adler}
  \emph {et~al.}}]{Adler:2003kt}%
  \BibitemOpen
  \bibfield  {author} {\bibinfo {author} {\bibfnamefont {S.~S.}\ \bibnamefont
  {Adler}} \emph {et~al.} (\bibinfo {collaboration} {PHENIX Collaboration}),\
  }\bibfield  {title} {\enquote {\bibinfo {title} {{Elliptic flow of identified
  hadrons in Au+Au collisions at $\sqrt{s_{NN}}$ = 200~GeV}},}\ }\href
  {\doibase 10.1103/PhysRevLett.91.182301} {\bibfield  {journal} {\bibinfo
  {journal} {Phys. Rev. Lett.}\ }\textbf {\bibinfo {volume} {91}},\ \bibinfo
  {pages} {182301} (\bibinfo {year} {2003})}\BibitemShut {NoStop}%
\bibitem [{\citenamefont {Bleicher}\ and\ \citenamefont
  {Stoecker}(2002)}]{Bleicher:2000sx}%
  \BibitemOpen
  \bibfield  {author} {\bibinfo {author} {\bibfnamefont {M.}~\bibnamefont
  {Bleicher}}\ and\ \bibinfo {author} {\bibfnamefont {H.}~\bibnamefont
  {Stoecker}},\ }\bibfield  {title} {\enquote {\bibinfo {title} {{Anisotropic
  flow in ultrarelativistic heavy ion collisions}},}\ }\href {\doibase
  10.1016/S0370-2693(01)01495-2} {\bibfield  {journal} {\bibinfo  {journal}
  {Phys. Lett. B}\ }\textbf {\bibinfo {volume} {526}},\ \bibinfo {pages} {309}
  (\bibinfo {year} {2002})}\BibitemShut {NoStop}%
\bibitem [{\citenamefont {Adare}\ \emph {et~al.}(2007)\citenamefont {Adare}
  \emph {et~al.}}]{Adare:2006ti}%
  \BibitemOpen
  \bibfield  {author} {\bibinfo {author} {\bibfnamefont {A.}~\bibnamefont
  {Adare}} \emph {et~al.} (\bibinfo {collaboration} {PHENIX Collaboration}),\
  }\bibfield  {title} {\enquote {\bibinfo {title} {{Scaling properties of
  azimuthal anisotropy in Au+Au and Cu+Cu collisions at $\sqrt{s}$ =
  200~GeV}},}\ }\href {\doibase 10.1103/PhysRevLett.98.162301} {\bibfield
  {journal} {\bibinfo  {journal} {Phys. Rev. Lett.}\ }\textbf {\bibinfo
  {volume} {98}},\ \bibinfo {pages} {162301} (\bibinfo {year}
  {2007})}\BibitemShut {NoStop}%
\bibitem [{\citenamefont {Adare}\ \emph {et~al.}(2012)\citenamefont {Adare}
  \emph {et~al.}}]{Adare:2012vq}%
  \BibitemOpen
  \bibfield  {author} {\bibinfo {author} {\bibfnamefont {A.}~\bibnamefont
  {Adare}} \emph {et~al.} (\bibinfo {collaboration} {PHENIX Collaboration}),\
  }\bibfield  {title} {\enquote {\bibinfo {title} {{Deviation from quark-number
  scaling of the anisotropy parameter $v_2$ of pions, kaons, and protons in
  Au+Au collisions at $\sqrt{s_{NN}}$=200 GeV}},}\ }\href {\doibase
  10.1103/PhysRevC.85.064914} {\bibfield  {journal} {\bibinfo  {journal} {Phys.
  Rev. C}\ }\textbf {\bibinfo {volume} {85}},\ \bibinfo {pages} {064914}
  (\bibinfo {year} {2012})}\BibitemShut {NoStop}%
\bibitem [{\citenamefont {Heinz}\ and\ \citenamefont
  {Snellings}(2013)}]{Heinz:2013th}%
  \BibitemOpen
  \bibfield  {author} {\bibinfo {author} {\bibfnamefont {U.}~\bibnamefont
  {Heinz}}\ and\ \bibinfo {author} {\bibfnamefont {R.}~\bibnamefont
  {Snellings}},\ }\bibfield  {title} {\enquote {\bibinfo {title} {{Collective
  flow and viscosity in relativistic heavy-ion collisions}},}\ }\href {\doibase
  10.1146/annurev-nucl-102212-170540} {\bibfield  {journal} {\bibinfo
  {journal} {Ann. Rev. Nucl. Part. Sci.}\ }\textbf {\bibinfo {volume} {63}},\
  \bibinfo {pages} {123} (\bibinfo {year} {2013})}\BibitemShut {NoStop}%
\bibitem [{\citenamefont {Ollitrault}(1992)}]{Ollitrault:1992bk}%
  \BibitemOpen
  \bibfield  {author} {\bibinfo {author} {\bibfnamefont {J.-Y.}\ \bibnamefont
  {Ollitrault}},\ }\bibfield  {title} {\enquote {\bibinfo {title} {{Anisotropy
  as a signature of transverse collective flow}},}\ }\href {\doibase
  10.1103/PhysRevD.46.229} {\bibfield  {journal} {\bibinfo  {journal} {Phys.
  Rev. D}\ }\textbf {\bibinfo {volume} {46}},\ \bibinfo {pages} {229} (\bibinfo
  {year} {1992})}\BibitemShut {NoStop}%
\bibitem [{\citenamefont {Niemi}\ \emph {et~al.}(2012)\citenamefont {Niemi},
  \citenamefont {Denicol}, \citenamefont {Huovinen}, \citenamefont {Molnar},\
  and\ \citenamefont {Rischke}}]{Niemi:2012ry}%
  \BibitemOpen
  \bibfield  {author} {\bibinfo {author} {\bibfnamefont {H.}~\bibnamefont
  {Niemi}}, \bibinfo {author} {\bibfnamefont {G.S.}\ \bibnamefont {Denicol}},
  \bibinfo {author} {\bibfnamefont {P.}~\bibnamefont {Huovinen}}, \bibinfo
  {author} {\bibfnamefont {E.}~\bibnamefont {Molnar}}, \ and\ \bibinfo {author}
  {\bibfnamefont {D.H.}\ \bibnamefont {Rischke}},\ }\bibfield  {title}
  {\enquote {\bibinfo {title} {{Influence of a temperature-dependent shear
  viscosity on the azimuthal asymmetries of transverse momentum spectra in
  ultrarelativistic heavy-ion collisions}},}\ }\href {\doibase
  10.1103/PhysRevC.86.014909} {\bibfield  {journal} {\bibinfo  {journal} {Phys.
  Rev. C}\ }\textbf {\bibinfo {volume} {86}},\ \bibinfo {pages} {014909}
  (\bibinfo {year} {2012})}\BibitemShut {NoStop}%
\bibitem [{\citenamefont {Song}\ \emph {et~al.}()\citenamefont {Song},
  \citenamefont {Bass}, \citenamefont {Heinz}, \citenamefont {Hirano},\ and\
  \citenamefont {Shen}}]{Song:2011hk}%
  \BibitemOpen
  \bibfield  {author} {\bibinfo {author} {\bibfnamefont {H.}~\bibnamefont
  {Song}}, \bibinfo {author} {\bibfnamefont {S.~A.}\ \bibnamefont {Bass}},
  \bibinfo {author} {\bibfnamefont {U.}~\bibnamefont {Heinz}}, \bibinfo
  {author} {\bibfnamefont {T.}~\bibnamefont {Hirano}}, \ and\ \bibinfo {author}
  {\bibfnamefont {C.}~\bibnamefont {Shen}},\ }\href@noop {} {\enquote {\bibinfo
  {title} {{Hadron spectra and elliptic flow for 200 A GeV Au+Au collisions
  from viscous hydrodynamics coupled to a Boltzmann cascade}},}\ }\bibinfo
  {note} {{Phys. Rev. C {\bf 83}, 054910 (2011); {\em Erratum: ibid} {\bf 86},
  059903E (2012)]}}\BibitemShut {NoStop}%
\bibitem [{\citenamefont {Soltz}\ \emph {et~al.}(2013)\citenamefont {Soltz},
  \citenamefont {Garishvili}, \citenamefont {Cheng}, \citenamefont {Abelev},
  \citenamefont {Glenn}, \citenamefont {Newby}, \citenamefont {Linden\~Levy},\
  and\ \citenamefont {Pratt}}]{Soltz:2012rk}%
  \BibitemOpen
  \bibfield  {author} {\bibinfo {author} {\bibfnamefont {R.~A.}\ \bibnamefont
  {Soltz}}, \bibinfo {author} {\bibfnamefont {I.}~\bibnamefont {Garishvili}},
  \bibinfo {author} {\bibfnamefont {M.}~\bibnamefont {Cheng}}, \bibinfo
  {author} {\bibfnamefont {B.}~\bibnamefont {Abelev}}, \bibinfo {author}
  {\bibfnamefont {A.}~\bibnamefont {Glenn}}, \bibinfo {author} {\bibfnamefont
  {J.~J.}\ \bibnamefont {Newby}}, \bibinfo {author} {\bibfnamefont {L.~A.}\
  \bibnamefont {Linden\~Levy}}, \ and\ \bibinfo {author} {\bibfnamefont
  {S.}~\bibnamefont {Pratt}},\ }\bibfield  {title} {\enquote {\bibinfo {title}
  {{Constraining the initial temperature and shear viscosity in a hybrid
  hydrodynamic model of $\sqrt{s_{NN}}$=200 GeV Au+Au collisions using pion
  spectra, elliptic flow, and femtoscopic radii}},}\ }\href {\doibase
  10.1103/PhysRevC.87.044901} {\bibfield  {journal} {\bibinfo  {journal} {Phys.
  Rev. C}\ }\textbf {\bibinfo {volume} {87}},\ \bibinfo {pages} {044901}
  (\bibinfo {year} {2013})}\BibitemShut {NoStop}%
\bibitem [{\citenamefont {Aamodt}\ \emph {et~al.}(2010)\citenamefont {Aamodt}
  \emph {et~al.}}]{Aamodt:2010pb}%
  \BibitemOpen
  \bibfield  {author} {\bibinfo {author} {\bibfnamefont {K}~\bibnamefont
  {Aamodt}} \emph {et~al.} (\bibinfo {collaboration} {ALICE Collaboration}),\
  }\bibfield  {title} {\enquote {\bibinfo {title} {{Charged-particle
  multiplicity density at mid-rapidity in central Pb-Pb collisions at
  $\sqrt{s_{NN}}$=2.76 TeV}},}\ }\href {\doibase
  10.1103/PhysRevLett.105.252301} {\bibfield  {journal} {\bibinfo  {journal}
  {Phys. Rev. Lett.}\ }\textbf {\bibinfo {volume} {105}},\ \bibinfo {pages}
  {252301} (\bibinfo {year} {2010})}\BibitemShut {NoStop}%
\bibitem [{\citenamefont {Aad}\ \emph {et~al.}(2012)\citenamefont {Aad} \emph
  {et~al.}}]{ATLAS:2012at}%
  \BibitemOpen
  \bibfield  {author} {\bibinfo {author} {\bibfnamefont {G.}~\bibnamefont
  {Aad}} \emph {et~al.} (\bibinfo {collaboration} {ATLAS Collaboration}),\
  }\bibfield  {title} {\enquote {\bibinfo {title} {{Measurement of the
  azimuthal anisotropy for charged particle production in $\sqrt{s_{NN}}$=2.76
  TeV lead-lead collisions with the ATLAS detector}},}\ }\href {\doibase
  10.1103/PhysRevC.86.014907} {\bibfield  {journal} {\bibinfo  {journal} {Phys.
  Rev. C}\ }\textbf {\bibinfo {volume} {86}},\ \bibinfo {pages} {014907}
  (\bibinfo {year} {2012})}\BibitemShut {NoStop}%
\bibitem [{\citenamefont {Chatrchyan}\ \emph
  {et~al.}(2013{\natexlab{a}})\citenamefont {Chatrchyan} \emph
  {et~al.}}]{Chatrchyan:2012ta}%
  \BibitemOpen
  \bibfield  {author} {\bibinfo {author} {\bibfnamefont {S.}~\bibnamefont
  {Chatrchyan}} \emph {et~al.} (\bibinfo {collaboration} {CMS Collaboration}),\
  }\bibfield  {title} {\enquote {\bibinfo {title} {{Measurement of the elliptic
  anisotropy of charged particles produced in PbPb collisions at
  $\sqrt{s}_{NN}$=2.76 TeV}},}\ }\href {\doibase 10.1103/PhysRevC.87.014902}
  {\bibfield  {journal} {\bibinfo  {journal} {Phys. Rev. C}\ }\textbf {\bibinfo
  {volume} {87}},\ \bibinfo {pages} {014902} (\bibinfo {year}
  {2013}{\natexlab{a}})}\BibitemShut {NoStop}%
\bibitem [{\citenamefont {Abelev}\ \emph {et~al.}(2013)\citenamefont {Abelev}
  \emph {et~al.}}]{Abelev:2012di}%
  \BibitemOpen
  \bibfield  {author} {\bibinfo {author} {\bibfnamefont {B.}~\bibnamefont
  {Abelev}} \emph {et~al.} (\bibinfo {collaboration} {ALICE Collaboration}),\
  }\bibfield  {title} {\enquote {\bibinfo {title} {{Anisotropic flow of charged
  hadrons, pions and (anti-)protons measured at high transverse momentum in
  Pb-Pb collisions at $\sqrt{s_{NN}}$=2.76 TeV}},}\ }\href {\doibase
  10.1016/j.physletb.2012.12.066} {\bibfield  {journal} {\bibinfo  {journal}
  {Phys. Lett. B}\ }\textbf {\bibinfo {volume} {719}},\ \bibinfo {pages} {18}
  (\bibinfo {year} {2013})}\BibitemShut {NoStop}%
\bibitem [{\citenamefont {Abelev}\ \emph {et~al.}()\citenamefont {Abelev} \emph
  {et~al.}}]{Abelev:2014pua}%
  \BibitemOpen
  \bibfield  {author} {\bibinfo {author} {\bibfnamefont {B.~B.}\ \bibnamefont
  {Abelev}} \emph {et~al.} (\bibinfo {collaboration} {ALICE Collaboration}),\
  }\href@noop {} {\enquote {\bibinfo {title} {{Elliptic flow of identified
  hadrons in Pb-Pb collisions at $\sqrt{s_{\rm{NN}}}$ = 2.76 TeV}},}\ }\bibinfo
  {note} {ArXiv:1405.4632}\BibitemShut {NoStop}%
\bibitem [{\citenamefont {Adamczyk}\ \emph {et~al.}(2013)\citenamefont
  {Adamczyk} \emph {et~al.}}]{Adamczyk:2013gw}%
  \BibitemOpen
  \bibfield  {author} {\bibinfo {author} {\bibfnamefont {L.}~\bibnamefont
  {Adamczyk}} \emph {et~al.} (\bibinfo {collaboration} {STAR Collaboration}),\
  }\bibfield  {title} {\enquote {\bibinfo {title} {{Elliptic flow of identified
  hadrons in Au+Au collisions at $\sqrt{s_{NN}}=$ 7.7-62.4 GeV}},}\ }\href
  {\doibase 10.1103/PhysRevC.88.014902} {\bibfield  {journal} {\bibinfo
  {journal} {Phys. Rev. C}\ }\textbf {\bibinfo {volume} {88}},\ \bibinfo
  {pages} {014902} (\bibinfo {year} {2013})}\BibitemShut {NoStop}%
\bibitem [{\citenamefont {Abelev}\ \emph {et~al.}(2010)\citenamefont {Abelev}
  \emph {et~al.}}]{Abelev:2010tr}%
  \BibitemOpen
  \bibfield  {author} {\bibinfo {author} {\bibfnamefont {B.~I.}\ \bibnamefont
  {Abelev}} \emph {et~al.} (\bibinfo {collaboration} {STAR Collaboration}),\
  }\bibfield  {title} {\enquote {\bibinfo {title} {{Charged and strange hadron
  elliptic flow in Cu+Cu collisions at $\sqrt{s_{NN}}$ = 62.4 and 200 GeV}},}\
  }\href {\doibase 10.1103/PhysRevC.81.044902} {\bibfield  {journal} {\bibinfo
  {journal} {Phys. Rev. C}\ }\textbf {\bibinfo {volume} {81}},\ \bibinfo
  {pages} {044902} (\bibinfo {year} {2010})}\BibitemShut {NoStop}%
\bibitem [{\citenamefont {Adare}\ \emph {et~al.}(2013)\citenamefont {Adare}
  \emph {et~al.}}]{Adare:2013piz}%
  \BibitemOpen
  \bibfield  {author} {\bibinfo {author} {\bibfnamefont {A.}~\bibnamefont
  {Adare}} \emph {et~al.} (\bibinfo {collaboration} {PHENIX Collaboration}),\
  }\bibfield  {title} {\enquote {\bibinfo {title} {{Quadrupole Anisotropy in
  Dihadron Azimuthal Correlations in Central $d$$+$Au Collisions at
  $\sqrt{s_{_{NN}}}$=200 GeV}},}\ }\href {\doibase
  10.1103/PhysRevLett.111.212301} {\bibfield  {journal} {\bibinfo  {journal}
  {Phys. Rev. Lett.}\ }\textbf {\bibinfo {volume} {111}},\ \bibinfo {pages}
  {212301} (\bibinfo {year} {2013})}\BibitemShut {NoStop}%
\bibitem [{\citenamefont {Khachatryan}\ \emph {et~al.}(2010)\citenamefont
  {Khachatryan} \emph {et~al.}}]{Khachatryan:2010gv}%
  \BibitemOpen
  \bibfield  {author} {\bibinfo {author} {\bibfnamefont {V.}~\bibnamefont
  {Khachatryan}} \emph {et~al.} (\bibinfo {collaboration} {CMS
  Collaboration}),\ }\bibfield  {title} {\enquote {\bibinfo {title}
  {{Observation of Long-Range Near-Side Angular Correlations in Proton-Proton
  Collisions at the LHC}},}\ }\href {\doibase 10.1007/J. High Energy
  Phys.09(2010)091} {\bibfield  {journal} {\bibinfo  {journal} {J. High Energy
  Phys.}\ }\textbf {\bibinfo {volume} {1009}},\ \bibinfo {pages} {091}
  (\bibinfo {year} {2010})}\BibitemShut {NoStop}%
\bibitem [{\citenamefont {Chatrchyan}\ \emph
  {et~al.}(2013{\natexlab{b}})\citenamefont {Chatrchyan} \emph
  {et~al.}}]{CMS:2012qk}%
  \BibitemOpen
  \bibfield  {author} {\bibinfo {author} {\bibfnamefont {S.}~\bibnamefont
  {Chatrchyan}} \emph {et~al.} (\bibinfo {collaboration} {CMS Collaboration}),\
  }\bibfield  {title} {\enquote {\bibinfo {title} {{Observation of long-range
  near-side angular correlations in proton-lead collisions at the LHC}},}\
  }\href {\doibase 10.1016/j.physletb.2012.11.025} {\bibfield  {journal}
  {\bibinfo  {journal} {Phys. Lett. B}\ }\textbf {\bibinfo {volume} {718}},\
  \bibinfo {pages} {795} (\bibinfo {year} {2013}{\natexlab{b}})}\BibitemShut
  {NoStop}%
\bibitem [{\citenamefont {Afanasiev}\ \emph
  {et~al.}(2009{\natexlab{a}})\citenamefont {Afanasiev} \emph
  {et~al.}}]{Afanasiev:2009wq}%
  \BibitemOpen
  \bibfield  {author} {\bibinfo {author} {\bibfnamefont {S.}~\bibnamefont
  {Afanasiev}} \emph {et~al.} (\bibinfo {collaboration} {PHENIX
  Collaboration}),\ }\bibfield  {title} {\enquote {\bibinfo {title}
  {{Systematic Studies of Elliptic Flow Measurements in Au+Au Collisions at
  $\sqrt{s}$ = 200~GeV}},}\ }\href {\doibase 10.1103/PhysRevC.80.024909}
  {\bibfield  {journal} {\bibinfo  {journal} {Phys. Rev. C}\ }\textbf {\bibinfo
  {volume} {80}},\ \bibinfo {pages} {024909} (\bibinfo {year}
  {2009}{\natexlab{a}})}\BibitemShut {NoStop}%
\bibitem [{\citenamefont {Adcox}\ \emph
  {et~al.}(2003{\natexlab{a}})\citenamefont {Adcox} \emph
  {et~al.}}]{Adcox:2003zm}%
  \BibitemOpen
  \bibfield  {author} {\bibinfo {author} {\bibfnamefont {K.}~\bibnamefont
  {Adcox}} \emph {et~al.} (\bibinfo {collaboration} {PHENIX Collaboration}),\
  }\bibfield  {title} {\enquote {\bibinfo {title} {{PHENIX detector
  overview}},}\ }\href {\doibase 10.1016/S0168-9002(02)01950-2} {\bibfield
  {journal} {\bibinfo  {journal} {Nucl. Instrum. Methods Phys. Res., Sec. A}\
  }\textbf {\bibinfo {volume} {499}},\ \bibinfo {pages} {469} (\bibinfo {year}
  {2003}{\natexlab{a}})}\BibitemShut {NoStop}%
\bibitem [{\citenamefont {Aizawa}\ \emph {et~al.}(2003)\citenamefont {Aizawa}
  \emph {et~al.}}]{Aizawa:2003zq}%
  \BibitemOpen
  \bibfield  {author} {\bibinfo {author} {\bibfnamefont {M.}~\bibnamefont
  {Aizawa}} \emph {et~al.} (\bibinfo {collaboration} {PHENIX Collaboration}),\
  }\bibfield  {title} {\enquote {\bibinfo {title} {{PHENIX central arm particle
  ID detectors}},}\ }\href {\doibase 10.1016/S0168-9002(02)01953-8} {\bibfield
  {journal} {\bibinfo  {journal} {Nucl. Instrum. Methods Phys. Res., Sec. A}\
  }\textbf {\bibinfo {volume} {499}},\ \bibinfo {pages} {508} (\bibinfo {year}
  {2003})}\BibitemShut {NoStop}%
\bibitem [{\citenamefont {Adcox}\ \emph
  {et~al.}(2003{\natexlab{b}})\citenamefont {Adcox} \emph
  {et~al.}}]{Adcox:2003en}%
  \BibitemOpen
  \bibfield  {author} {\bibinfo {author} {\bibfnamefont {K.}~\bibnamefont
  {Adcox}} \emph {et~al.} (\bibinfo {collaboration} {PHENIX Collaboration}),\
  }\bibfield  {title} {\enquote {\bibinfo {title} {{Construction and
  performance of the PHENIX pad chambers}},}\ }\href {\doibase
  10.1016/S0168-9002(02)01791-6} {\bibfield  {journal} {\bibinfo  {journal}
  {Nucl. Instrum. Methods Phys. Res., Sec. A}\ }\textbf {\bibinfo {volume}
  {497}},\ \bibinfo {pages} {263} (\bibinfo {year}
  {2003}{\natexlab{b}})}\BibitemShut {NoStop}%
\bibitem [{\citenamefont {Aphecetche}\ \emph {et~al.}(2003)\citenamefont
  {Aphecetche} \emph {et~al.}}]{Aphecetche:2003zr}%
  \BibitemOpen
  \bibfield  {author} {\bibinfo {author} {\bibfnamefont {L.}~\bibnamefont
  {Aphecetche}} \emph {et~al.} (\bibinfo {collaboration} {PHENIX
  Collaboration}),\ }\bibfield  {title} {\enquote {\bibinfo {title} {{PHENIX
  calorimeter}},}\ }\href {\doibase 10.1016/S0168-9002(02)01954-X} {\bibfield
  {journal} {\bibinfo  {journal} {Nucl. Instrum. Methods Phys. Res., Sec. A}\
  }\textbf {\bibinfo {volume} {499}},\ \bibinfo {pages} {521} (\bibinfo {year}
  {2003})}\BibitemShut {NoStop}%
\bibitem [{\citenamefont {Adler}\ \emph {et~al.}(2005)\citenamefont {Adler}
  \emph {et~al.}}]{Adler:2004zn}%
  \BibitemOpen
  \bibfield  {author} {\bibinfo {author} {\bibfnamefont {S.~S.}\ \bibnamefont
  {Adler}} \emph {et~al.} (\bibinfo {collaboration} {PHENIX Collaboration}),\
  }\bibfield  {title} {\enquote {\bibinfo {title} {{Systematic studies of the
  centrality and $\sqrt{s_{NN}}$ dependence of the d E(T) / d eta and d (N(ch)
  / d eta in heavy ion collisions at mid-rapidity}},}\ }\href {\doibase
  10.1103/PhysRevC.71.049901, 10.1103/PhysRevC.71.034908} {\bibfield  {journal}
  {\bibinfo  {journal} {Phys. Rev. C}\ }\textbf {\bibinfo {volume} {71}},\
  \bibinfo {pages} {034908} (\bibinfo {year} {2005})}\BibitemShut {NoStop}%
\bibitem [{\citenamefont {Alver}\ \emph {et~al.}(2007)\citenamefont {Alver}
  \emph {et~al.}}]{Alver:2006wh}%
  \BibitemOpen
  \bibfield  {author} {\bibinfo {author} {\bibfnamefont {B.}~\bibnamefont
  {Alver}} \emph {et~al.} (\bibinfo {collaboration} {PHOBOS Collaboration}),\
  }\bibfield  {title} {\enquote {\bibinfo {title} {{System size, energy,
  pseudorapidity, and centrality dependence of elliptic flow}},}\ }\href
  {\doibase 10.1103/PhysRevLett.98.242302} {\bibfield  {journal} {\bibinfo
  {journal} {Phys. Rev. Lett.}\ }\textbf {\bibinfo {volume} {98}},\ \bibinfo
  {pages} {242302} (\bibinfo {year} {2007})}\BibitemShut {NoStop}%
\bibitem [{\citenamefont {Miller}\ \emph {et~al.}(2007)\citenamefont {Miller},
  \citenamefont {Reygers}, \citenamefont {Sanders},\ and\ \citenamefont
  {Steinberg}}]{Miller:2007ri}%
  \BibitemOpen
  \bibfield  {author} {\bibinfo {author} {\bibfnamefont {M.~L.}\ \bibnamefont
  {Miller}}, \bibinfo {author} {\bibfnamefont {K.}~\bibnamefont {Reygers}},
  \bibinfo {author} {\bibfnamefont {Stephen~J.}\ \bibnamefont {Sanders}}, \
  and\ \bibinfo {author} {\bibfnamefont {Peter}\ \bibnamefont {Steinberg}},\
  }\bibfield  {title} {\enquote {\bibinfo {title} {{Glauber modeling in high
  energy nuclear collisions}},}\ }\href {\doibase
  10.1146/annurev.nucl.57.090506.123020} {\bibfield  {journal} {\bibinfo
  {journal} {Ann. Rev. Nucl. Part. Sci.}\ }\textbf {\bibinfo {volume} {57}},\
  \bibinfo {pages} {205} (\bibinfo {year} {2007})}\BibitemShut {NoStop}%
\bibitem [{\citenamefont {Mitchell}\ \emph {et~al.}(2002)\citenamefont
  {Mitchell} \emph {et~al.}}]{Mitchell:2002wu}%
  \BibitemOpen
  \bibfield  {author} {\bibinfo {author} {\bibfnamefont {J.~T.}\ \bibnamefont
  {Mitchell}} \emph {et~al.} (\bibinfo {collaboration} {PHENIX
  Collaboration}),\ }\bibfield  {title} {\enquote {\bibinfo {title} {{Event
  reconstruction in the PHENIX central arm spectrometers}},}\ }\href {\doibase
  10.1016/S0168-9002(01)01512-1} {\bibfield  {journal} {\bibinfo  {journal}
  {Nucl. Instrum. Methods Phys. Res., Sec. A}\ }\textbf {\bibinfo {volume}
  {482}},\ \bibinfo {pages} {491} (\bibinfo {year} {2002})}\BibitemShut
  {NoStop}%
\bibitem [{\citenamefont {Adler}\ \emph {et~al.}(2004)\citenamefont {Adler}
  \emph {et~al.}}]{Adler:2003au}%
  \BibitemOpen
  \bibfield  {author} {\bibinfo {author} {\bibfnamefont {S.~S.}\ \bibnamefont
  {Adler}} \emph {et~al.} (\bibinfo {collaboration} {PHENIX Collaboration}),\
  }\bibfield  {title} {\enquote {\bibinfo {title} {{High $p_{T}$ charged hadron
  suppression in Au + Au collisions at $\sqrt{s}_{NN} = 200$ GeV}},}\ }\href
  {\doibase 10.1103/PhysRevC.69.034910} {\bibfield  {journal} {\bibinfo
  {journal} {Phys. Rev. C}\ }\textbf {\bibinfo {volume} {69}},\ \bibinfo
  {pages} {034910} (\bibinfo {year} {2004})}\BibitemShut {NoStop}%
\bibitem [{\citenamefont {Adler}\ \emph {et~al.}(2006)\citenamefont {Adler}
  \emph {et~al.}}]{Adler:2005ad}%
  \BibitemOpen
  \bibfield  {author} {\bibinfo {author} {\bibfnamefont {S.~S.}\ \bibnamefont
  {Adler}} \emph {et~al.} (\bibinfo {collaboration} {PHENIX Collaboration}),\
  }\bibfield  {title} {\enquote {\bibinfo {title} {{Jet structure from dihadron
  correlations in d+Au collisions at $\sqrt{s_{NN}}$ = 200~GeV}},}\ }\href
  {\doibase 10.1103/PhysRevC.73.054903} {\bibfield  {journal} {\bibinfo
  {journal} {Phys. Rev. C}\ }\textbf {\bibinfo {volume} {73}},\ \bibinfo
  {pages} {054903} (\bibinfo {year} {2006})}\BibitemShut {NoStop}%
\bibitem [{\citenamefont {Poskanzer}\ and\ \citenamefont
  {Voloshin}(1998)}]{Poskanzer:1998yz}%
  \BibitemOpen
  \bibfield  {author} {\bibinfo {author} {\bibfnamefont {A.~M.}\ \bibnamefont
  {Poskanzer}}\ and\ \bibinfo {author} {\bibfnamefont {S.~A.}\ \bibnamefont
  {Voloshin}},\ }\bibfield  {title} {\enquote {\bibinfo {title} {{Methods for
  analyzing anisotropic flow in relativistic nuclear collisions}},}\ }\href
  {\doibase 10.1103/PhysRevC.58.1671} {\bibfield  {journal} {\bibinfo
  {journal} {Phys. Rev. C}\ }\textbf {\bibinfo {volume} {58}},\ \bibinfo
  {pages} {1671} (\bibinfo {year} {1998})}\BibitemShut {NoStop}%
\bibitem [{\citenamefont {Adare}\ \emph {et~al.}(2010)\citenamefont {Adare}
  \emph {et~al.}}]{Adare:2010ux}%
  \BibitemOpen
  \bibfield  {author} {\bibinfo {author} {\bibfnamefont {A.}~\bibnamefont
  {Adare}} \emph {et~al.} (\bibinfo {collaboration} {PHENIX Collaboration}),\
  }\bibfield  {title} {\enquote {\bibinfo {title} {{Elliptic and hexadecapole
  flow of charged hadrons in Au+Au collisions at $\sqrt{s_{NN}}$=200 GeV}},}\
  }\href {\doibase 10.1103/PhysRevLett.105.062301} {\bibfield  {journal}
  {\bibinfo  {journal} {Phys. Rev. Lett.}\ }\textbf {\bibinfo {volume} {105}},\
  \bibinfo {pages} {062301} (\bibinfo {year} {2010})}\BibitemShut {NoStop}%
\bibitem [{\citenamefont {Afanasiev}\ \emph
  {et~al.}(2009{\natexlab{b}})\citenamefont {Afanasiev} \emph
  {et~al.}}]{Afanasiev:2009ii}%
  \BibitemOpen
  \bibfield  {author} {\bibinfo {author} {\bibfnamefont {S.}~\bibnamefont
  {Afanasiev}} \emph {et~al.} (\bibinfo {collaboration} {PHENIX
  Collaboration}),\ }\bibfield  {title} {\enquote {\bibinfo {title} {{Kaon
  interferometric probes of space-time evolution in Au+Au collisions at
  $\sqrt{s_{NN}}$ = 200~GeV}},}\ }\href {\doibase
  10.1103/PhysRevLett.103.142301} {\bibfield  {journal} {\bibinfo  {journal}
  {Phys. Rev. Lett.}\ }\textbf {\bibinfo {volume} {103}},\ \bibinfo {pages}
  {142301} (\bibinfo {year} {2009}{\natexlab{b}})}\BibitemShut {NoStop}%
\bibitem [{\citenamefont {Adler}\ \emph {et~al.}(2002)\citenamefont {Adler}
  \emph {et~al.}}]{Adler:2002pu}%
  \BibitemOpen
  \bibfield  {author} {\bibinfo {author} {\bibfnamefont {C.}~\bibnamefont
  {Adler}} \emph {et~al.} (\bibinfo {collaboration} {STAR Collaboration}),\
  }\bibfield  {title} {\enquote {\bibinfo {title} {{Elliptic flow from two and
  four particle correlations in Au+Au collisions at $\sqrt{s_{NN}}$ =
  130~GeV}},}\ }\href {\doibase 10.1103/PhysRevC.66.034904} {\bibfield
  {journal} {\bibinfo  {journal} {Phys. Rev. C}\ }\textbf {\bibinfo {volume}
  {66}},\ \bibinfo {pages} {034904} (\bibinfo {year} {2002})}\BibitemShut
  {NoStop}%
\bibitem [{\citenamefont {Abelev}\ \emph {et~al.}(2007)\citenamefont {Abelev}
  \emph {et~al.}}]{Abelev:2007qg}%
  \BibitemOpen
  \bibfield  {author} {\bibinfo {author} {\bibfnamefont {B.~I.}\ \bibnamefont
  {Abelev}} \emph {et~al.} (\bibinfo {collaboration} {STAR Collaboration}),\
  }\bibfield  {title} {\enquote {\bibinfo {title} {{Mass, quark-number, and
  $\sqrt{s_{NN}}$ dependence of the second and fourth flow harmonics in
  ultra-relativistic nucleus-nucleus collisions}},}\ }\href {\doibase
  10.1103/PhysRevC.75.054906} {\bibfield  {journal} {\bibinfo  {journal} {Phys.
  Rev. C}\ }\textbf {\bibinfo {volume} {75}},\ \bibinfo {pages} {054906}
  (\bibinfo {year} {2007})}\BibitemShut {NoStop}%
\bibitem [{\citenamefont {Voloshin}\ \emph {et~al.}(2008)\citenamefont
  {Voloshin}, \citenamefont {Poskanzer},\ and\ \citenamefont
  {Snellings}}]{Voloshin:2008dg}%
  \BibitemOpen
  \bibfield  {author} {\bibinfo {author} {\bibfnamefont {S.~A.}\ \bibnamefont
  {Voloshin}}, \bibinfo {author} {\bibfnamefont {A.~M.}\ \bibnamefont
  {Poskanzer}}, \ and\ \bibinfo {author} {\bibfnamefont {R.}~\bibnamefont
  {Snellings}},\ }\href@noop {} {\enquote {\bibinfo {title} {{Collective
  phenomena in non-central nuclear collisions}},}\ } (\bibinfo {year} {2008}),\
  \bibinfo {note} {arXiv:0809.2949}\BibitemShut {NoStop}%
\bibitem [{\citenamefont {Borghini}(2006)}]{Borghini:2005hx}%
  \BibitemOpen
  \bibfield  {author} {\bibinfo {author} {\bibfnamefont {N.}~\bibnamefont
  {Borghini}},\ }\bibfield  {title} {\enquote {\bibinfo {title} {{Hints of
  incomplete thermalization in RHIC data}},}\ }\href {\doibase
  10.1140/epja/i2005-10293-9} {\bibfield  {journal} {\bibinfo  {journal} {Eur.
  Phys. J. A}\ }\textbf {\bibinfo {volume} {29}},\ \bibinfo {pages} {27}
  (\bibinfo {year} {2006})}\BibitemShut {NoStop}%
\bibitem [{\citenamefont {Uphoff}\ \emph {et~al.}(2015)\citenamefont {Uphoff},
  \citenamefont {Senzel}, \citenamefont {Fochler}, \citenamefont {Wesp},
  \citenamefont {Xu} \emph {et~al.}}]{Uphoff:2014cba}%
  \BibitemOpen
  \bibfield  {author} {\bibinfo {author} {\bibfnamefont {J.}~\bibnamefont
  {Uphoff}}, \bibinfo {author} {\bibfnamefont {F.}~\bibnamefont {Senzel}},
  \bibinfo {author} {\bibfnamefont {O.}~\bibnamefont {Fochler}}, \bibinfo
  {author} {\bibfnamefont {C.}~\bibnamefont {Wesp}}, \bibinfo {author}
  {\bibfnamefont {Z.}~\bibnamefont {Xu}},  \emph {et~al.},\ }\bibfield  {title}
  {\enquote {\bibinfo {title} {{Elliptic flow and nuclear modification factor
  in ultrarelativistic heavy-ion collisions within a partonic transport
  model}},}\ }\href {\doibase 10.1103/PhysRevLett.114.112301} {\bibfield
  {journal} {\bibinfo  {journal} {Phys. Rev. Lett.}\ }\textbf {\bibinfo
  {volume} {114}},\ \bibinfo {pages} {112301} (\bibinfo {year}
  {2015})}\BibitemShut {NoStop}%
\bibitem [{\citenamefont {Adare}\ \emph {et~al.}()\citenamefont {Adare} \emph
  {et~al.}}]{Adare:2014kci}%
  \BibitemOpen
  \bibfield  {author} {\bibinfo {author} {\bibfnamefont {A.}~\bibnamefont
  {Adare}} \emph {et~al.} (\bibinfo {collaboration} {PHENIX Collaboration}),\
  }\href@noop {} {\enquote {\bibinfo {title} {{Measurement of the higher-order
  anisotropic flow coefficients for identified hadrons in Au$+$Au collisions at
  $\sqrt{s_{_{NN}}}$ = 200 GeV}},}\ }\bibinfo {note}
  {ArXiv:1412.1038}\BibitemShut {NoStop}%
\end{thebibliography}

%
 
\end{document}